\documentclass[10pt]{amsproc}
\usepackage{xcolor}
\usepackage{graphicx,hyperref}
\usepackage[normalem]{ulem}
\usepackage{amsfonts,amssymb,amsmath,amsgen,amsthm}

\newcommand{\msc}[2][2000]{%
  \let\@oldtitle\@title%
  \gdef\@title{\@oldtitle\footnotetext{#1 \emph{Mathematics subject
        classification.} #2}}% 
}

\def\({\left(}
\def\){\right)}
\def\<{\left\langle}
\def\>{\right\rangle}
\def\le{\leqslant}

\begin{document}

%%%% Article title to be placed here
\title[Dynamical approximations for composite quantum systems]{Dynamical approximations for composite quantum systems:
Assessment of error estimates for a separable ansatz}

\author[I. Burghardt]{Irene Burghardt}
\address{Institute of Physical \& Theoretical Chemistry, Goethe University Frankfurt, 
%Max-von-Laue-Str. 7, 60438 Frankfurt am Main, 
Germany }
%\ead{burghardt@chemie.uni-frankfurt.de}
\email{burghardt@chemie.uni-frankfurt.de}

\author[R. Carles]{R\'emi Carles}
\address{Univ Rennes, CNRS, IRMAR - UMR 6625,  F-35000 Rennes, France}
%\ead{Remi.Carles@math.cnrs.fr}
\email{Remi.Carles@math.cnrs.fr}

\author[C. Fermanian]{Clotilde Fermanian Kammerer}
\address{
Univ Paris Est Cr\'eteil, CNRS, LAMA, F-94010 Cr\'eteil, France\\ 
Univ Gustave Eiffel, LAMA, F-77447 Marne-la-Vallée, France}
%\ead{clotilde.fermanian@u-pec.fr}
\email{clotilde.fermanian@u-pec.fr}

\author[B. Lasorne]{Benjamin Lasorne*}
\address{ICGM, Univ Montpellier, CNRS, ENSCM, Montpellier, France}
%\ead{benjamin.lasorne@umontpellier.fr}
\email{benjamin.lasorne@umontpellier.fr}

\author[C. Lasser]{Caroline Lasser*}
\address{Technische Universit\"at M\"unchen, Zentrum Mathematik, Deutschland}
%\ead{classer@ma.tum.de}
\email{classer@ma.tum.de}

\begin{abstract}
Numerical studies are presented to assess error estimates for a
separable (Hartree) approximation for dynamically evolving composite quantum
systems which exhibit distinct scales defined by their mass and frequency
ratios. The relevant error estimates were formally described in our previous
work [I. Burghardt, R. Carles, C. Fermanian Kammerer, B. Lasorne, C. Lasser, J. Phys.
A. 54, 414002 (2021)]. Specifically, we consider a representative
two-dimensional tunneling system where a double well and a harmonic coordinate
are cubically coupled. The time-dependent Hartree approximation is compared
with a fully correlated solution, for different parameter regimes. The
impact of the coupling and the resulting correlations are quantitatively
assessed in terms of a time-dependent reaction probability along the tunneling
coordinate. We show that the numerical error is correctly predicted on
moderate time scales by a theoretically derived error estimate.
\end{abstract}

\thanks{* Corresponding authors.}
\maketitle

%%%%%%%%%%%%%%%%%%%%%%%%%%%%%%%%%%%%

\section{Introduction}

The time-dependent Hartree (TDH) approximation, also termed
time-dependent self-consistent field method
\cite{Dirac1930,Gerber1982,Gerber1988}, which represents the time propagation
of composite quantum systems within a separable (Hartree) approximation, is
ubiquitous in quantum and classical statistical physics. This approximation is
based on a mean-field description and often works well if the relevant
subspaces are weakly coupled, and if a separation of scales is given due to
disparities in masses and/or frequencies. The TDH approximation is also a natural
starting point for including correlations in terms of sums of products, i.e.,
using a correlated multiconfigurational (MC) ansatz that leads to a
Multiconfiguration Time-Dependent Hartree (MCTDH) \cite{Meyer90,BECK20001} form of the
wavefunction. Related tensor representations of multidimensional wavefunctions are cast in
the form of matrix product states \cite{Schro_Chin,Strat}. 
A variational setting \cite{BECK20001,LubichBlue} is generally employed to obtain
generalized, multiconfigurational mean-field equations for such correlated
wavefunctions. The TDH and MCTDH representations can be straightforwardly
adapted to fermionic or bosonic systems. In the present context, we refer to
distinguishable particles for simplicity.

Despite the importance of the TDH ansatz, an explicit error analysis
of this approach is not often reported in the literature. In a recent
formal paper \cite{BCFLL}, we therefore presented error estimates for
the time propagation of composite quantum systems within the TDH
approximation. We also compared different types of approximate product
wavefunctions, i.e., based on Taylor expansion (collocation) or else
on the TDH mean-field approach, and we further considered a
semiclassical approximation within a quantum-classical type
treatment. Such semiclassical, or quantum-classical
approximations are especially useful if the system is composite
-- or structured -- in a physical sense such that the subsystems
exhibit different time and/or energy scales.
In the present paper, we follow up on this previous work
and carry out numerical simulations to assess the previously derived
error estimates for a realistic, anharmonically coupled system
exhibiting a separation of scales defined by the relevant mass and
frequency ratios. As in the formal paper mentioned above, the present
study is meant to be a first step towards a general analysis of scale
separation in the context of multiconfigurational, tensorized
wavefunction representations. 

Specifically, we consider numerical simulations for a two-dimensional
tunneling system where a double-well potential is anharmonically
coupled to a harmonic coordinate. As in Ref. \cite{BCFLL}, a cubic
coupling is considered (i.e., linear in the tunneling coordinate and
quadratic in the harmonic coordinate). Numerical TDH calculations for
different parameter regimes are compared with correlated MCTDH
calculations that can be considered as numerically exact for the
present system. The impact of the coupling and the resulting
correlations are quantitatively assessed in terms of a time-dependent
reaction probability along the tunneling coordinate. 

A time-dependent error estimate is expressed quantitatively in terms
of the relevant parameters of the Hamiltonian, leading to insight into
the ``small'' parameters to be considered to gauge the validity of the
TDH approximation. This leads us to question the conventional
viewpoint that mass ratios are decisive; indeed, within our present
analysis, it is the frequency ratio that is found to play a more
important role. 

 The present study paves the way for a more general
treatment including the effects of fluctuations and dissipation,
which are expected to have a non-trivial effect on the tunneling
dynamics \cite{Weiss:12,Breuer:02}. 

The paper is organized as follows. Section 2 presents the model Hamiltonian under consideration and gives a detailed account of the relevant parameters and their scaling properties. Section 3 addresses the TDH approximation and Section 4 details the formulation of time-dependent error estimates. Section 5 presents the numerical results and Section 6 concludes.

\section{Model system}
In line with our previous work \cite{BCFLL}, we consider a two-dimensional
model system which is of system-bath type and exhibits anharmonicities both
within the system subspace and in the system-bath coupling. Specifically, a
double-well potential is chosen in the system subspace, which is coupled to a
harmonic bath coordinate via a cubic (quadratic times linear) coupling term.
As pointed out in our previous analysis \cite{BCFLL}, a cubic coupling is a
non-trivial case which is relevant for the description of vibrational
dephasing \cite{Levine:88,Gruebele:04} and Fermi resonances \cite{Bunker:06}
in a molecular physics context.

Here and in the following, we adhere to a system-bath
terminology despite the low-dimensional nature of the model system, due to the
fact that the low-frequency harmonic vibration coupled to the dominant
tunneling coordinate essentially acts as a bath mode. From a more rigorous
viewpoint, the single bath coordinate under consideration carries
non-Markovian effects that emerge from a decomposition of memory effects in
terms of effective-mode chains \cite{Martinazzo:11,Martinazzo2:11}. Following this
description, we recently considered a related two-dimensional tunneling system
where a single effective bath mode is augmented by a Markovian master equation
\cite{Picconi19}. Likewise, the present treatment can be augmented such as to
yield a full system-bath treatment. However, the purpose of the present work
is to investigate dynamical approximations for the effective bath mode that is
accounted for explicitly.

From a complementary viewpoint, the present system can be
considered as a typical example of multi-dimensional tunneling which is
frequently encountered in polyatomic molecular systems \cite{Nakamura} and has
been extensively investigated, both experimentally and theoretically
\cite{Shida:89,Wassermann:06,Woo:19,Qu:21}. Multi-dimensional tunneling
involves multiple time scales, resonance effects, vibrational mode
selectivity, and non-statistical energy redistribution. The present system,
even though comparatively simple, falls into a typical parameter regime of
such molecular systems, and will be shown to exhibit some of the features
mentioned above, specifically the observation of multiple time scales and the
interference of seemingly passive spectator modes with the tunneling process.

\subsection{Hamiltonian in physical scaling}

In system-bath form, the Hamiltonian is written as follows,
\begin{equation}
{\mathcal H} = H_X + H_Y + W(X,Y),
\nonumber
\end{equation}
with the subspace Hamiltonians
\begin{equation}
H_X = -\frac{\hbar^2}{2\mu_1}\Delta_X + V_1(X), \qquad\qquad
H_Y = -\frac{\hbar^2}{2\mu_2}\Delta_Y + V_2(Y),
\nonumber
\end{equation}
where $V_1(X)$ corresponds to a double-well potential
and $V_2(Y)$ is a harmonic form, 
\begin{equation}
V_1(X) = \frac12 k^0_1 X^2 \left(\frac{X}{2L}-1\right)^2,
\qquad\qquad V_2(Y) = \frac12 k^0_2 Y^2.
\nonumber
\end{equation}
The system potential $V_1(X)$ represents a symmetric double well
with two equivalent minima at $X=0$ and $X=2L$ that can be taken to
correspond to the ``reactant'' vs.\ ``product'' in the context of
a reactive process \cite{Nakamura}. The energy at the barrier $X=L$ amounts
to $D_1 = k_1^0L^2/8$.

The coupling $W(X,Y)$ is given as a cubic term, i.e., linear in $X$ and
quadratic in~$Y$,
\begin{equation}
W(X,Y) = \frac12 \eta^0 XY^2.
\nonumber
\end{equation}
Viewed from a different angle, $V_2(Y)$ and $W(X,Y)$ can be combined
into an effective potential for the $Y$ coordinate,
\[
V_2^{\rm eff} = V_2(Y) + W(X,Y) = \frac{1}{2} \left( k^0_2 + \eta^0 X \right)
Y^2 \equiv \frac{1}{2} k_2(X) Y^2,
\]
whose curvature is $X$-dependent, i.e., the local force constant is
given as $k_2(X) = k_2^0 + \eta^0 X$. We denote the ratio
of the curvature along $Y$, taken at the product versus reactant minima in
$X$, by
\begin{equation}\label{alpha}
\alpha = \frac{k_2(X=2L)}{k_2(X=0)} = 1 + 2 \frac{\eta^0}{k_2^0}L.
\end{equation}
The reduced parameter $\alpha$ gives a direct measure of the relative
coupling strength upon characterizing how much the local curvature for $Y$
changes as $X$ varies from $0$ (reactant minimum) to $2L$ (product minimum).
 
Yet from a different perspective, an adiabatic regime can be considered
whose ``fast'' subsystem ($X$) coordinate is coupled to a ``slow'' bath 
($Y$) coordinate. An effective subsystem Hamiltonian can then be defined
as follows,
\[
H_1^{\rm eff} = -\frac{\hbar^2}{2\mu_1}\Delta_X
+ V(X,Y) \qquad
V(X, Y) = V_1(X) + V_2(Y) + W(X, Y) .
\]
Which perspective is most appropriate depends on the physical time scales
under study, as will be detailed below. 

Finally, we note that the extension of this model to multivariate coordinates
$X$ and~$Y$ is straightforward, such that the present model is suitable
to address general multidimensional tunneling situations.

\subsection{Parameter choice}
In the present work, we exclusively consider negative
values of the coupling parameter $\eta^0$  such that the
curvature ratio defined in Eq.~(\ref{alpha}) satisfies $\alpha < 1$.
This ensures that the ground eigenstate of the
full $(X,Y)$-system  is localized around the product well
($X=2L>0$), while nonstationary dynamics will start from an
initial condition (or, in mathematical language, an initial
datum) localized around the reactant well ($X=0$). In other words,
we reverse localization in order to create an initial
nonstationary state.

The cubic coupling potential $W(X,Y)$ has to be handled with some care. It
causes the potential energy to be unbounded from below beyond the controllable
subquadratic regime. This is likely to induce numerical issues and renders the
Hamiltonian ${\mathcal H}$ only formally self-adjoint. One might therefore
either add a contribution to the potential energy that is quartically
confining with respect to $Y$ or multiply the coupling potential with a
cut-off function in~$X$. Here we have chosen another simple approach: There
exists a critical value $X=X_\mathrm{c}=\frac{2L}{1-\alpha} > 0$ where
$k_2(X=X_\mathrm{c})=0$, which is called a valley-ridge inflection point
\cite{LasVRI03}. Despite its
relevance in terms of bifurcation aspects, this is
not the situation that we want to address here. A simple cure is to ensure
that such a critical point occurs far enough from $X=2L$ that the potential
energy at this point,
$V(X=X_\mathrm{c},Y=0)=V_1(X_\mathrm{c})$, is large enough compared to the
barrier height, $D_1 = k_1^0L^2/8$. As will be shown below, we choose our
reference model such that $\alpha=\frac13$ and $X_\mathrm{c}=3L$, which
implies that $V_1(X_\mathrm{c})$ is nine times larger than $D_1$. The most
``precarious'' case we considered is $\alpha=\frac14$ and
$X_\mathrm{c}=\frac{8L}{3}$, which implies that we have $V_1(X_\mathrm{c}) =
\frac{256}{81} D_1 \approx 3.2 D_1$. Such a range of values for the onset of
unboundedness in the potential energy seems \emph{a priori} far enough that
our low-energy wavepackets will be vanishing in critical regions, which is
what we also observed in practice.

\subsection{Representation in scaled coordinates}
In order to transform the Hamiltonian to a suitably scaled
representation, we introduce the (angular) frequencies and the corresponding natural length
scales of the harmonic approximations
for $X$ and~$Y$ around the origin,
\[
\omega_i^0 = \sqrt{k_i^0/\mu_i},\quad
a_i^0 = \sqrt{\hbar/(\mu_i\omega_i^0)}
\]
for $i=1,2$. The corresponding natural energy and time scales are
\[
E_i^0 = \frac{\hbar^2}{\mu_i (a_i^0)^2} = k_i^0 (a_i^0)^2 = \hbar\omega_i^0,\quad
t_i^0 = \frac{\hbar}{E_i^0,}.
\]
Note that, \emph{e.g.}, $\{\mu_1,a_1^0,t_1^0\}$ can serve as a consistent and
complete set of mechanical units for all quantities built on powers of
[M][L][T], whereby $\hbar$ is numerically equal to unity due to the
relationship between the energy unit $E_1^0$ and the time unit~$t_1^0$ (much as
when considering atomic units). In the spirit of semiclassical scaling, we
further introduce a typical parameter defined as the square root of the mass
ratios,
\[
\varepsilon = \sqrt{\frac{\mu_1}{\mu_2}}.
\]
Scaling both coordinates with respect 
to the natural length scale of the 
system coordinate, while the bath coordinate is additionally scaled
by $\varepsilon$, 
we set
\[
(x,y) = \left(\frac{1}{a_1^0} X,\frac{1}{a_1^0 \varepsilon} Y\right),\quad \ell = \frac{L}{a_1^0}.
\]
We thus obtain a scaled Hamiltonian
\[
\mathcal{H} = E_1^0  H,
\]
where the dimensionless part reads
\[
H = -\frac{1}{2}\Delta_x - \frac{1}{2}\Delta_y + v(x,y),
\]
with potential energy
\[
v(x,y) = v_1(x) + v_2(y) + w(x,y),
\]
with contributions from the system (double well)
and the harmonic bath mode
\[
v_1(x) = \frac{1}{2} x^2\left(\frac{x}{2\ell}-1\right)^2,\quad
v_2(y) = \frac{1}{2}\varpi^2 y^2,  
\]
where $\varpi = \omega_2^0/\omega_1^0$ denotes the frequency ratio of the bath versus the system. The coupling potential
$w(x,y) = \frac12 \eta xy^2$  features
the rescaled coupling constant 
\[
\eta = \frac{(a_1^0)^3 \varepsilon^2}{E_1^0}
\,\eta^0.
\]
Let us denote the physical time $T$ and the rescaled time $t$, where
\[
t = \frac{T}{t_1^0} = \frac{E_1^0}{\hbar}T.
\]
We can absorb both $E_1^0$ and $\hbar$ into $t_1^0$ and recast the time-dependent Schr\"odinger equation, 
\[
i\hbar \partial_T\Psi(T,X,Y) = \mathcal{H} \Psi(T,X,Y),
\]
into dimensionless form as
\[
 i\partial_t\psi(t,x,y) = H \psi(t,x,y).
\]
We observe that the curvature ratio defined in Eq.~(\ref{alpha}) is invariant under the performed linear coordinate scaling. 
It is related to the coupling constant according to
\[
\alpha = 1 + 2 \frac{\eta}{\varpi^2}\ell
\quad\mbox{resp.}\quad
\eta = \varpi^2 \frac{\alpha-1}{2\ell}.
\]
The parametrization in terms of the frequency ratio
$\varpi$ and the curvature ratio $\alpha$
fully characterizes the interaction of the system
and bath via the potential energy.
The parameter $\varepsilon$, which represents the
mass ratio, only appears indirectly, through the scaling of the
$Y$ coordinate and of the $\eta$ parameter.

\subsection{Relevant parameter regime and initial data}

In view of the numerical simulation results reported below, we now
specify the parameter regime which was considered in these simulations. First,
a reference model was constructed by the following choice of parameters,
\[
\varepsilon_*= \frac14,\ \varpi_*=\frac{1}{100},\ \alpha_* = \frac13.
\]
 We note that the mass ratio $\varepsilon_*$ is moderate,
but representative of chemically relevant systems. The frequency ratio
$\varpi_*$, however, takes a value that is significantly smaller, and will be demonstrated to play
an important role in the error estimates to be discussed below.

In the simulations reported below,
we explore the dynamics of several variations
of this reference model  (i.e., cases 0 to 8, while the
reference model is denoted case *, see Table~\ref{tab:cases}). Relevant ranges of values
for $\alpha$ have been discussed above. As already pointed out,
reducing $\alpha$ below
values of about $\frac14$ could entail issues related to the unboundedness within
the space explored by the wavepacket. Note also that ``case 3''
appeared in our simulations as the most sensitive situation, bringing much
larger errors between correlated and uncorrelated descriptions. We suspect
that this may reflect the fact that the corresponding time scale of the $Y$ dynamics, now
shorter, enters the realm of the time scale of $X$ motion (see Sec. \ref{sec:res-disc}
for further discussion), thus making system-bath separability less justified.

\begin{figure}\centering
{\includegraphics[width = 0.7\textwidth]{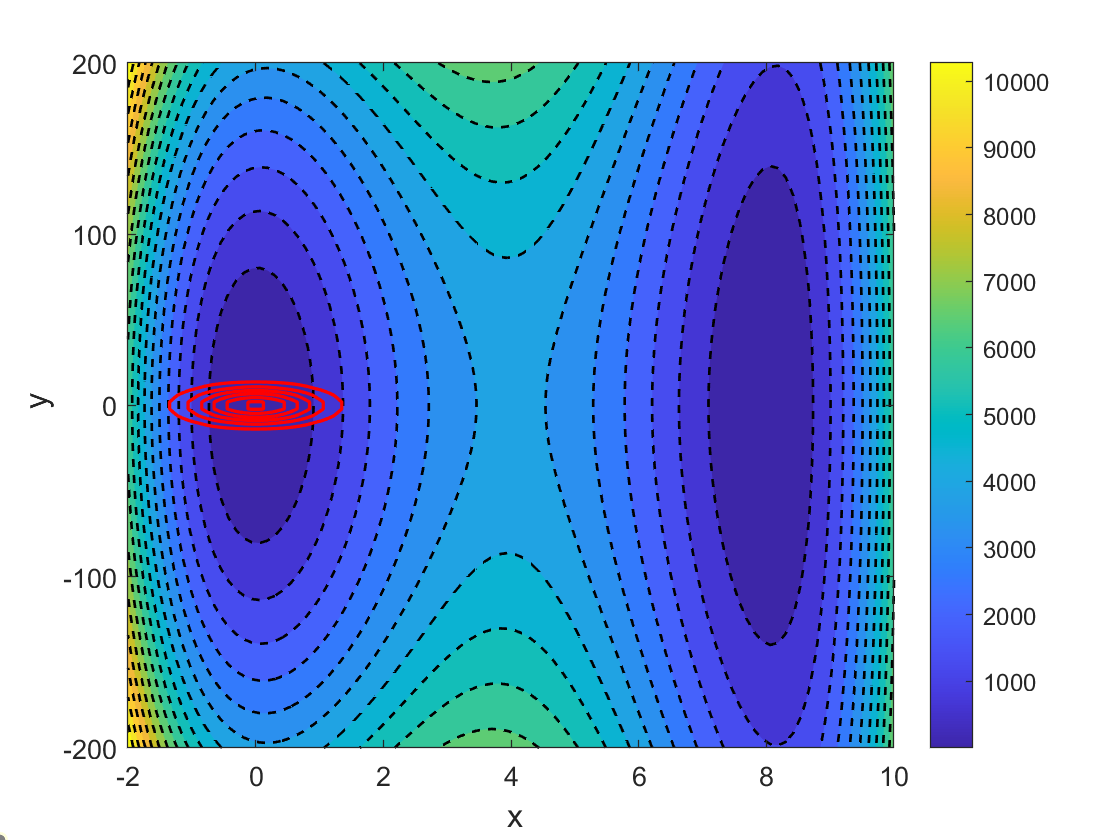}}\\
{\includegraphics[width = 0.7\textwidth, height = 0.3\textheight]{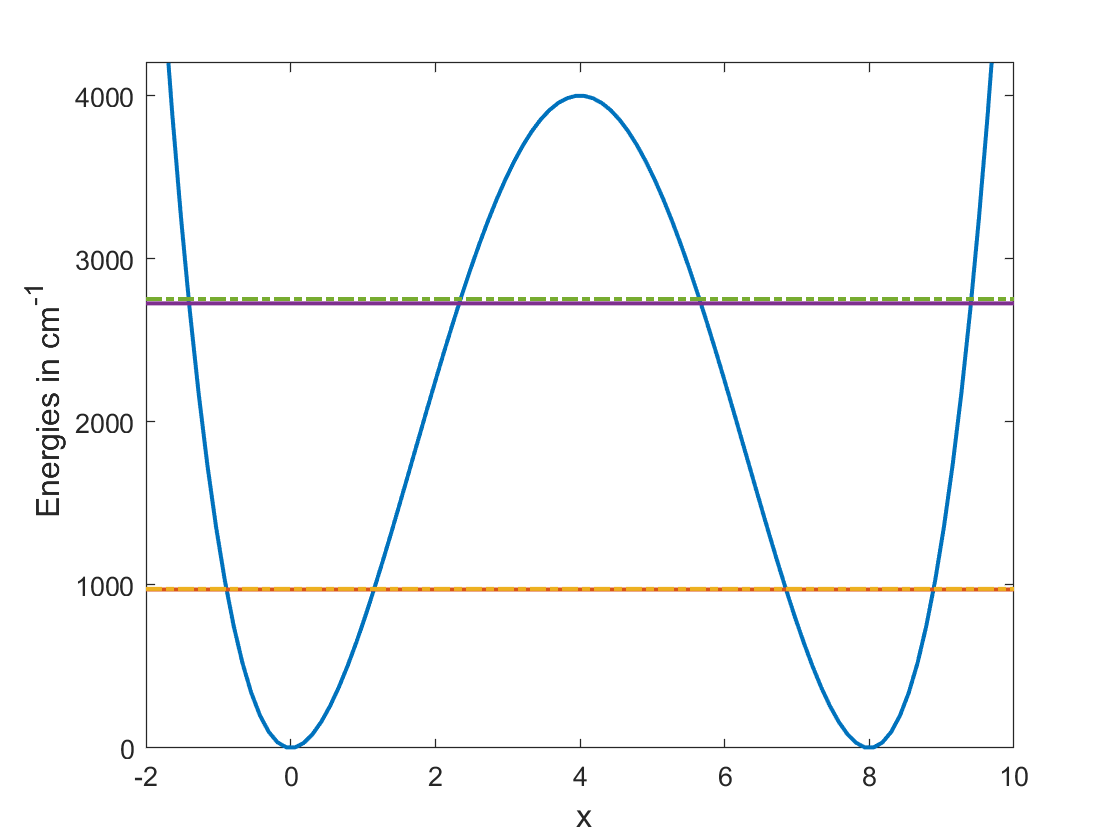}}
\caption{Upper panel: Contour plot of the two-dimensional potential energy surface (reference model *)
and coherent-state initial condition according to Eq.~(\ref{initial-cond}). As
explained in the text, the initial condition corresponds to a non-stationary
state localized in the less stable left (``reactant'') well. Lower panel: One-dimensional 
cut at $y=0$ through the potential and eigenenergies of the ground-state and 
first excited-state tunneling pairs. The energies are $972$ and $976$~cm$^{-1}$ for the ground pair, 
$2726$ and $2753$~cm$^{-1}$ for the excited pair.  
The splitting of the even and odd energy levels is barely visible on the energy scale set by the potential.}
\label{figure1}
\end{figure}

In all cases, we chose the initial data to be the approximate Gaussian
quasi-coherent ground state localized in the ``reactant well",
 as illustrated in Fig.~\ref{figure1},
\begin{equation}
\psi(t=0,x,y) = \chi_0(x)\phi_0(y) = (2\pi)^{-1/2}\varpi^{-1/4}\exp\left(
-\frac{1}{2}x^2 - \frac{\varpi}{2}y^2\right).
\label{initial-cond}
\end{equation}
We note that both minima of the potential energy have equal energies, but the zero-point energy is higher in the left reactant well because the curvature for $y$ is higher there.
The squeezing in the $y$ direction, due to the $\varpi$ factor in the coherent
state width, reflects that this coordinate tends
to the classical limit.

\begin{table}[h]\centering
\begin{tabular}{|c|c|c|c|l|}
\hline
Case & $\varepsilon$ & $\varpi$ & $\alpha$ & \\\hline
* & $\varepsilon_*$ & $\varpi_*$ & $\alpha_*$ & reference\\\hline
0 & $\varepsilon_*$ & $\varpi_*$ & $1$ & no coupling\\\hline
1 & $2\varepsilon_*$ & $\varpi_*$ & $\alpha_*$ & \\
2 & $\frac12\varepsilon_*$ & $\varpi_*$ & $\alpha_*$ & \\\hline
3 & $\varepsilon_*$ & $4\varpi_*$ & $\alpha_*$ & most sensitive\\
4 & $\varepsilon_*$ & $\frac14\varpi_*$ & $\alpha_*$ & \\\hline
5 & $\varepsilon_*$ & $\varpi_*$ & $\frac34\alpha_*$ & \\
6 & $\varepsilon_*$ & $\varpi_*$ & $\frac32\alpha_*$ & \\
7 & $\varepsilon_*$ & $\varpi_*$ & $2\alpha_*$ & \\
8 & $\varepsilon_*$ & $\varpi_*$ & $\frac94\alpha_*$ & \\
\hline 
\end{tabular}
\caption{Parameter variations of the reference model, that is defined by the values $(\varepsilon_*,\varpi_*,\alpha_*)=(\frac14,\frac{1}{100},\frac13)$. These parameters determine the square root of the mass ratio, the frequency ratio, and the curvature ratio, respectively.}
\label{tab:cases}
\end{table}

\section{Time-dependent Hartree approximation}
We compare the  numerical solution of the full Schr\"odinger equation with separable, normalized initial datum
\begin{eqnarray*}
\left\{
\begin{array}{l}
i\partial_t\psi(t,x,y) =  H\psi(t,x,y),\\
\psi(t=0,x,y)=\chi_0(x)\phi_0(y) ,
\end{array}
\right.
\end{eqnarray*}
with the one of the  TDH approximation subject to the same initial condition, 
\begin{eqnarray*}
\left\{
\begin{array}{l} i\partial_t u(t,x,y) = H^{\rm eff}_u(t)u(t,x,y),\\
 u(t=0,x,y)=\chi_0(x)\phi_0(y).
 \end{array}
\right.
\end{eqnarray*}
The  effective TDH Hamiltonian 
\[
H^{\rm eff}_u(t) 
= \langle H\rangle_{\phi}^{(x)}(t) + 
\langle H\rangle_{\chi}^{(y)}(t) - \langle H\rangle_{u}(t),
\]
is additive with respect to the coordinates and thus preserves the product
structure of the initial
datum,  i.e., we have
\[
u(t,x,y) = a(t)\,\chi(t,x)\,\phi(t,y),
\]
where $a(t)$ is a complex number of absolute value one, acting as a suitably chosen gauge factor. 
The individual product wavefunctions satisfy the coupled equations of motion
\begin{eqnarray*}
i\partial_t \chi(t,x) = \langle  H\rangle_\phi(t)\chi(t,x),\\
i\partial_t \phi(t,y) = \langle  H\rangle_{\chi}(t)   
\phi(t,y),
\end{eqnarray*}
while the gauge factor 
\[
a(t) = \exp\left(i\int_0^t \langle H\rangle_u(s\,) d s\right)
\]
depends on the full energy expectation with respect to the Hartree product. Our system-bath type Hamiltonian is of the form 
\[
H =  H_x +  H_y + w(x,y),
\]
with $H_x = -\frac12\Delta_x + v_1(x)$ and 
$ H_y = -\frac12\Delta_y + v_2(y)$. In this situation, the effective Hartree Hamiltonian is of the form
\[
H^{\rm eff}_u(t) =  H_x +  H_y + w_u^{\rm eff}(t,x,y),
\]
and differs from the true Hamiltonian only with respect to the effective coupling potential
\begin{equation}\label{wapp}
w^{\rm eff}_u(t,x,y) = \langle w\rangle_\phi(t,x) + 
\langle w\rangle_\chi(t,y) - \langle w\rangle_u(t),
\end{equation}
so the above effective Hamiltonians can be rewritten as
\[
    \langle H\rangle_\phi = H_x + \langle w\rangle_\phi(t,x) ,\quad \langle H\rangle_{\chi} = H_y + \langle w\rangle_\chi(t,y) .
\]

\section{Error estimates}

For analyzing the approximation error
\[
e(t,x,y) = \psi(t,x,y) - u(t,x,y),
\]
we use a standard stability estimate, see Lemma~2 in Ref.\ \cite{BCFLL} Differentiating the error with respect to time we obtain a Schr\"odinger-type equation
\begin{eqnarray*}
\left\{
\begin{array}{l}
i\partial_t e(t,x,y) = H e(t,x,y) + \Sigma_u(t,x,y),\\
e(t=0,x,y) = 0,
\end{array}
\right.
\end{eqnarray*}
with source term
\begin{eqnarray*}
\Sigma_u(t,x,y) &= \left(H - H^{\rm eff}_u(t) \right)u(t,x,y)= \left(w(x,y) - w^{\rm eff}_u(t,x,y) \right)u(t,x,y).
\end{eqnarray*}
By the variation of constants formula (aka the Duhamel principle), we write the error as a time-integral, 
\begin{equation}\label{error}
e(t,x,y) = \frac{1}{i}\int_0^t \exp(-i H(t-s)) \Sigma_u(s,x,y) \,d s.
\end{equation}
Since the time-evolution associated with the Hamiltonian 
$H$ is unitary, we now estimate
\[
\|e(t)\|_{L^2} \le \int_0^t \|\Sigma_u(s)\|_{L^2} \, ds
\]
for the $L^2$-norm of the error.

\subsection{Formula for the source term}
The cubic coupling potential of our system-bath type Hamiltonian is of product form
\[
w(x,y)=w_1(x)w_2(y).
\]
Therefore, the coupling potential of the Hartree approximation, see Eq.~(\ref{wapp}), takes the special form
\[
w^{\rm eff}_u(t,x,y) = w_1(x)\,\langle w_2\rangle_{\phi}(t)+ 
\langle w_1\rangle_{\chi}(t)\, w_2(y) 
- \langle w_1\rangle_{\chi}(t)\, \langle w_2\rangle_{\phi}(t).
\]
This implies for the difference of the coupling potentials
\begin{eqnarray*}
\delta w_u(t,x,y) 
&= w(x,y) - w^{\rm eff}_u(t,x,y)  \\
&=\left(w_1(x)-\langle w_1\rangle_{\chi}(t)\right)
\left(w_2(y) - \langle w_2\rangle_{\phi}(t)\right).
\end{eqnarray*}
We provide a detailed computation of the local-in-time error given in
Example~3 of Ref.\ \cite{BCFLL}. 
We calculate the norm of the source term $\Sigma_u(t) = \delta w_u(t)u(t)$ according to
\begin{eqnarray*}
\|\Sigma_u(t)\|_{L^2}^2 &= 
\langle\delta w^2\rangle_u(t)\\
&=\left\langle \left[w_1(x)-\langle w_1\rangle_{\chi}(t)\right]^2\right\rangle_\chi\, 
\left\langle \left[w_2(y) - \langle w_2\rangle_{\phi}(t)\right]^2\right\rangle_\phi\\
&=\left(\langle w_1^2\rangle_\chi(t) - \langle w_1\rangle_\chi(t)^2\right)
\left(\langle w_2^2\rangle_\phi(t) - \langle w_2\rangle_\phi(t)^2\right).
\end{eqnarray*}
From a probabilistic point of view, we can interpret this formula as the product of the variances of $w_1$ and $w_2$. 
Applying this formula to the cubic coupling model, we then obtain the error estimate
\[
\|\psi(t)-u(t)\|_{L^2} \le \frac12|\eta| \int_0^t 
\sqrt{\left(\langle x^2\rangle_\chi(s)-\langle x\rangle_{\chi}^2(s)\right)
\left( \langle y^4\rangle_\phi(s) - \langle y^2\rangle_{\phi}^2(s)\right)}\ d s.
\]

\subsection{Dimension analysis}
This formula is given here in terms of dimensionless energy, space, and time variables; its proof is not affected by scaling considerations, and it directly translates into physical units, $(X,Y,T)$, as
\begin{eqnarray}
&\|\Psi(T)-U(T)\|_{L^2} \le \frac12|\eta^0| \times\nonumber \\*[1ex]
&\frac{1}{\hbar} \int_0^T 
\sqrt{\left(\langle X^2\rangle_X(S)-\langle X\rangle_{X}^2(S)\right)
\left(\langle Y^4\rangle_Y(S) - \langle Y^2\rangle_{Y}^2(S)\right)} \ d S.
\label{eq:error-estimate}
\end{eqnarray}
From $\eta^0 = \frac{E_1^0}{(a_1^0)^3 \varepsilon^2}\,\eta$, recalling $E_1^0 = \frac{\hbar}{t_1^0,}$, and $\eta=\varpi^2\varsigma$, where 
\[
\varsigma= \frac{\alpha-1}{2\ell},
\]
the upper bound of the previous estimate can be recast in terms of dimensionless ratios as
\begin{eqnarray*}
&\frac{\varpi^2|\varsigma|}{2t_1^0} \int_0^T 
\sqrt{\frac{\langle X^2\rangle_X(S)-\langle X\rangle_{X}^2(S)}{(a_1^0)^2}
\frac{\langle Y^4\rangle_Y(S) - \langle Y^2\rangle_{Y}^2(S)}{\epsilon^4(a_1^0)^4}} \ d S\\*[1ex]
&=\frac{\varpi^2|\varsigma|}{2t_1^0} \int_0^T 
\sqrt{\frac{\langle X^2\rangle_X(S)-\langle X\rangle_{X}^2(S)}{(a_1^0)^2}
\frac{\langle Y^4\rangle_Y(S) - \langle Y^2\rangle_{Y}^2(S)}{\varpi^2(a_2^0)^4}} \ d S\\*[1ex]
&=\frac{\varpi|\varsigma|}{2t_1^0} \int_0^T 
\sqrt{\frac{\langle X^2\rangle_X(S)-\langle X\rangle_{X}^2(S)}{(a_1^0)^2}
\frac{\langle Y^4\rangle_Y(S) - \langle Y^2\rangle_{Y}^2(S)}{(a_2^0)^4}} \ d S,
\end{eqnarray*}
where we used
\[
\frac{a_2^0}{a_1^0}=\frac{\varepsilon}{\sqrt{\varpi}},
\]
and eliminated the somewhat artificial dependence on $\varepsilon$ (noting
that different
values of $\varepsilon$ only bring homothetic dynamics with respect to $Y$). As a crucial
consequence, we removed one power order of $\varpi$ regarding natural orders
of magnitude and effective ``smallness''.

\subsection{Linearization of the upper bound}
From an operational point of view, the purpose of rescaling essentially
consists in determining relevant orders of magnitude for the values of the
various factors entering  the relevant formulae. Since we specifically chose initial data to
be quasi-coherent states (within a harmonic approximation around the origin),
we know that the product of $X$ and $Y^2$ standard deviations expressed in
their respective natural units satisfy
\[
\sqrt{\frac{\langle X^2\rangle_X(T=0)-\langle X\rangle_{X}^2(T=0)}{(a_1^0)^2}}\times 
\sqrt{\frac{\langle Y^4\rangle_Y(T=0) - \langle Y^2\rangle_{Y}^2(T=0)}{(a_2^0)^4}}
=\frac12,
\]
and will not change dramatically over time, which is the incentive for considering a rescaling based on natural units. We can thus further propose a sort of ``rough'' linear estimate for short times as follows,
\begin{equation}
\|\Psi(T)-U(T)\|_{L^2} \lesssim \frac14\varpi|\varsigma| \frac{T}{t_1^0} .
\label{eq:linear-approx}
\end{equation}
where $\lesssim$ here is to be understood as preceding an approximate upper bound. Such an approximation is not aimed at being precise beyond very short times (although we shall see later on that it works surprisingly well at later times) but it presents the great advantage of providing an easy estimate of relevant orders of magnitude before performing any actual propagation. In the present situation, the initial widths along $X$ and $Y$ were chosen to vary as little as possible over time (quasi-coherent initial datum); however, it is not so difficult to make a rough prediction of the time evolution of standard deviations in more general cases (especially oscillatory breathing behaviors with harmonic half periods).

It is worth noticing that we have identified the prefactor $\varpi \varsigma =
({\eta}/{\varpi})$ that appears in Eq.~(\ref{eq:linear-approx}) as
an objective measure of the impact of the coupling on the rate of growth of the
error with respect to time. This was not evident at first sight when starting
from $\eta^0$ as in Eq.~(\ref{eq:error-estimate}) written with physical units.
It required the dimension analysis presented above so as to get rid of
dimensioned quantities and identify what will take values close to unity. The
parameter $\varsigma$ only affects the coupling between the system and the
bath but can only be varied moderately. In contrast, $\varpi$ can span a large
range; however, changing its value affects both the coupling and the relative
timescales between system and bath.

We also  emphasize that the error  in the norm of the difference between
two normalized wavefunctions is limited  by a strict
upper bound, a ``maximum maximorum'', which is $\sqrt{2}$ (in the worst and
undesired case of strict orthogonality between the solution and its
approximation). The intersection of our estimate with this value gives a
maximal time of relevance for the estimate, but also a predictive rough order
of magnitude of the time beyond which an uncorrelated approximation is
definitely at risk.

\section{Results and discussion}\label{sec:res-disc}
\begin{figure}\centering
{\includegraphics[width = 0.7\textwidth]{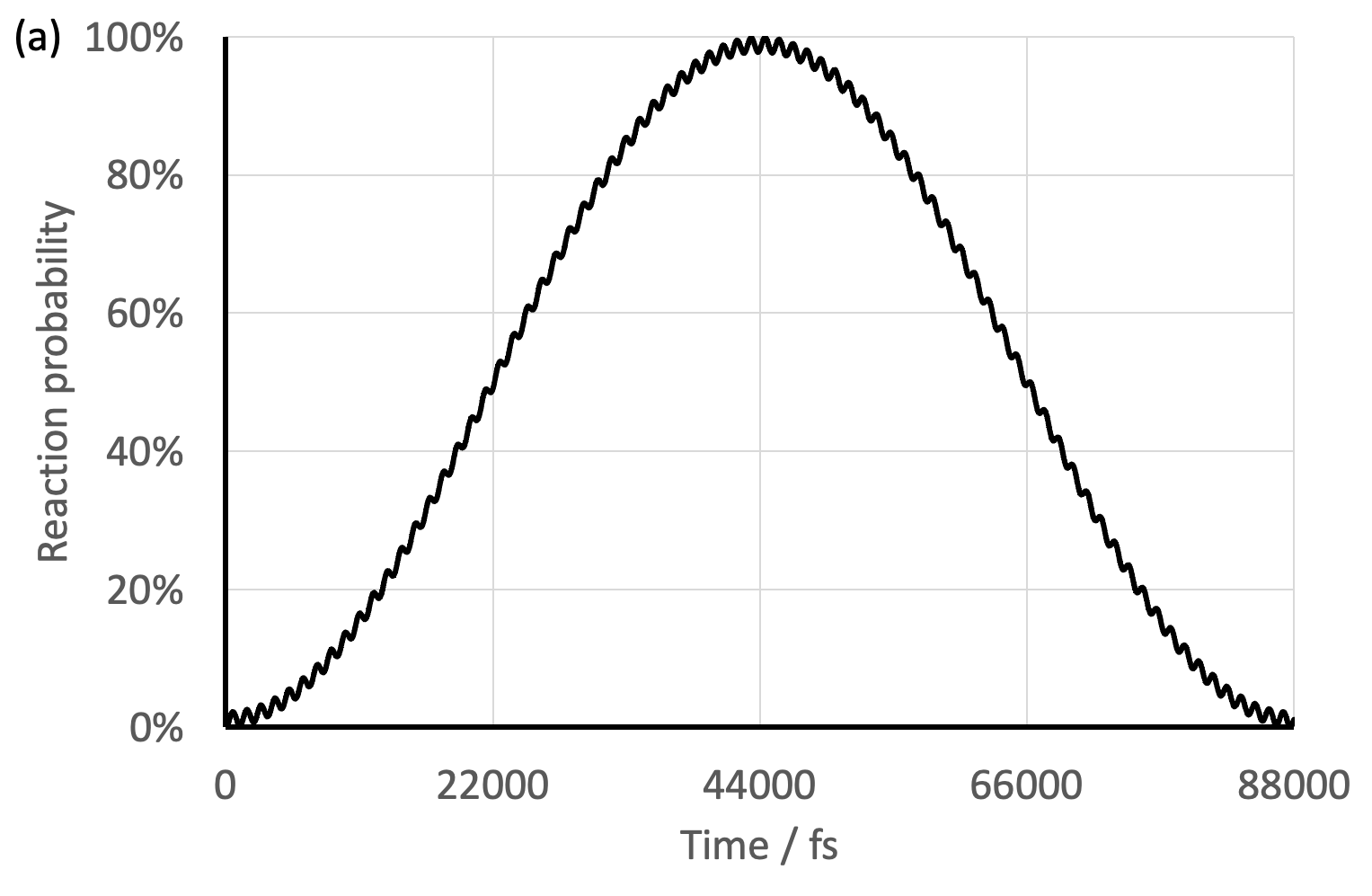}}\\
{\includegraphics[width = 0.7\textwidth]{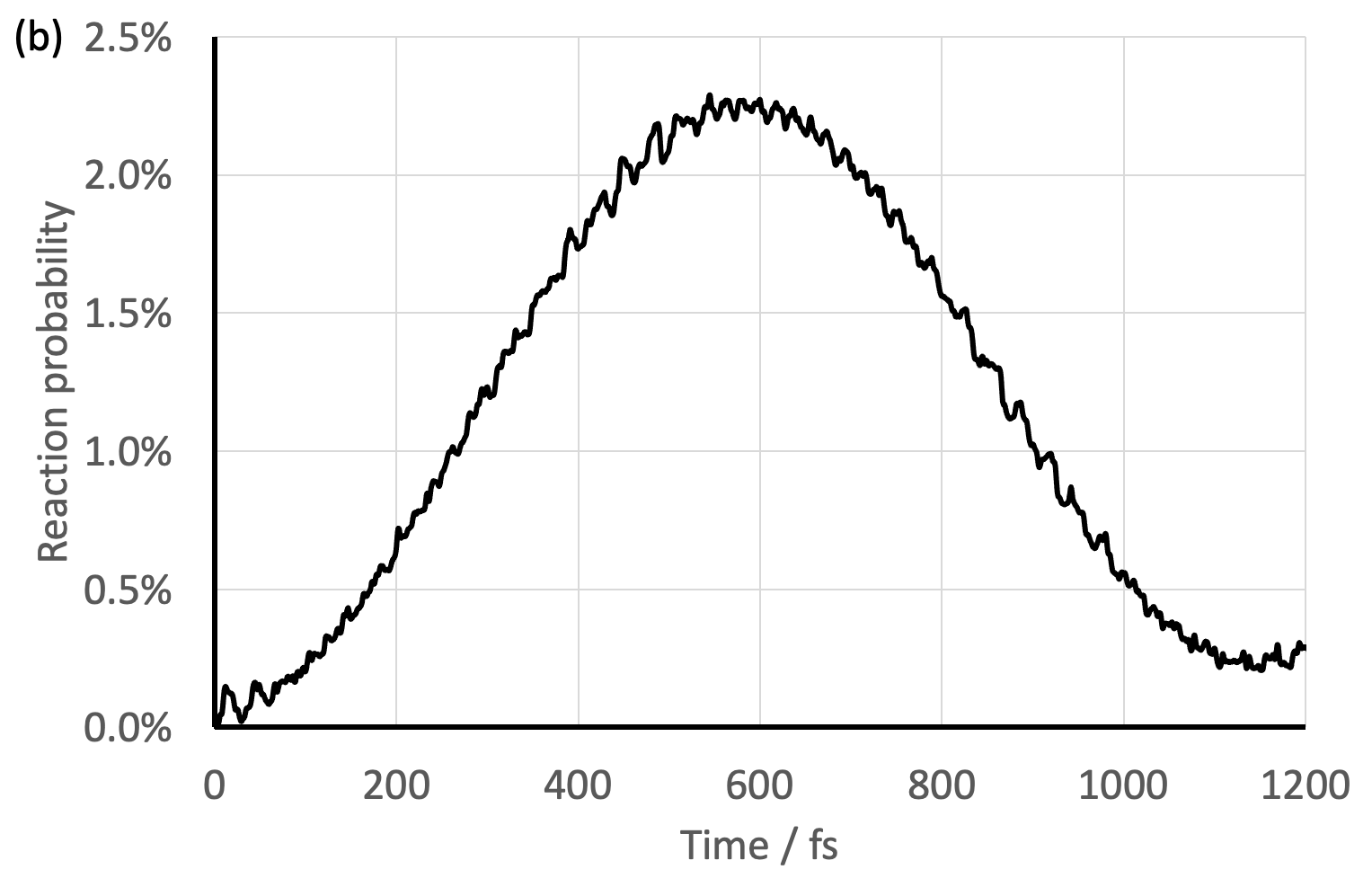}}\\
{\includegraphics[width = 0.7\textwidth]{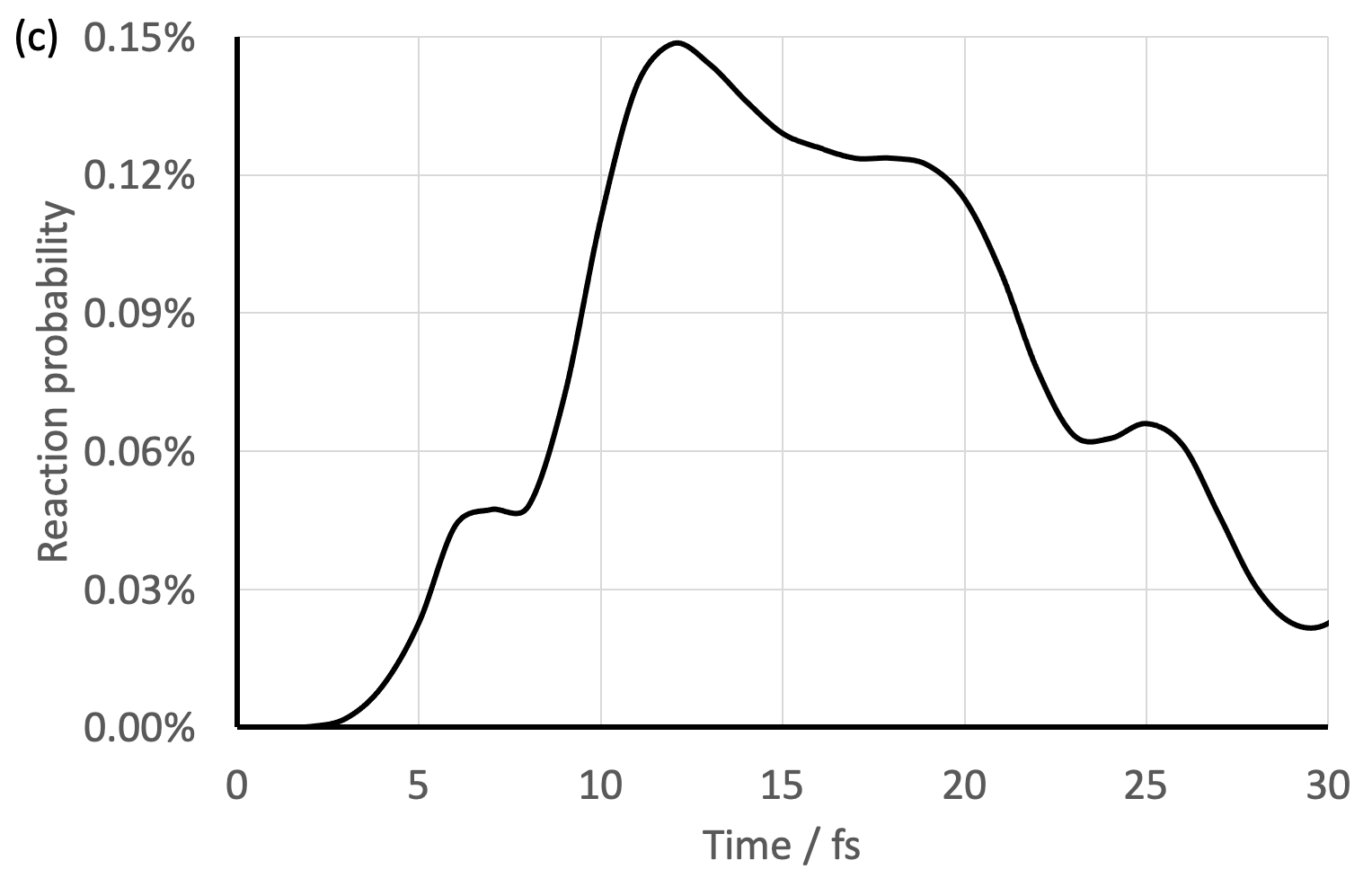}}
\caption{ Fully correlated (MCTDH) propagation: Time evolution of the reaction probability,
$R(T)$, in the uncoupled model (case 0) ; (a) long times (ground-state
tunneling); (b) medium times (excited-state tunneling); (c) short times (quasiharmonic).}
\label{fig:FigA1}
\end{figure}
All simulations presented below were computed with the Quantics software
\cite{wor2020_quantics}.  Reference simulations
denoted as ``fully correlated'' in the following refer to converged MCTDH calculations where correlated system-bath
states are propagated under variational equations of motion \cite{Meyer90,BECK20001,LubichBlue}. In the
specific case of two degrees of freedom, the MCTDH wavefunction ansatz reads as
follows, as a generalization of the TDH ansatz,
\begin{equation}
\psi(t, x, y) =  \sum_{j_1=1}^{n} \sum_{j_2=1}^{n} A_{j_1j_2}(t)
\varphi_{j_1}(t,x) \chi_{j_2}(t,y)
\end{equation}  
The convergence of the multiconfigurational expansion is measured in terms of
the so-called natural weights (natural orbital populations) \cite{BECK20001}. Typically, expansions up to
$n=3$ single-particle functions (orbitals) were necessary in order to achieve convergence for the present systems.
We achieved numerically converged situations over long times with natural weights of about 99 \%, 1 \%, and less than 0.1 \% for both degrees of freedom.
Further computational details and convergence analysis are
provided in the Supplementary Material [LINK].
Related MCTDH calculations for two-dimensional tunneling situations are
reported, e.g., in Ref.\ \cite{Picconi19}. 

\subsection{Characteristic times of dynamical simulations}\label{sec:characteristic}

For reference, we first consider the uncoupled model (case $0$; see
Table \ref{tab:cases}). The characteristics of our 
model are such that we can distinguish three very different time scales, all
separated by two orders of magnitude, for the $X$ subsystem dynamics: long,
short, and medium times. The long time scale $\sim$100 ps
relates to ``ground-state tunneling'' (induced by the energy
splitting between the ground-state tunneling pair); the medium
time scale $\sim$ 1 ps is ``excited-state
tunneling'' (first excited tunneling pair splitting); the short time $\sim$
0.01 ps is due to ``quasiharmonic'' motion
(vibration around the local minimum).

Our preferred observable  for monitoring the dynamics will be the
``reaction probability'', $R(T) = \langle H_L\rangle(T)$. It is defined as the
expectation value of the Heaviside step distribution centered at $X=L$, such
that it provides a measure of the probability for the system to be in the
product region ($X>L$) at any given time.

In the uncoupled case, the time evolution of $R(T)$ shows a perfect tunneling
quantum beat between the ``reactant" ($X=0$) and ``product'' ($X=2L$) wells. It
oscillates between 0 and 1 with a long period of about 88 ps; see
Fig.~\ref{fig:FigA1}. There is a clear modulation (around 2\%) with a medium
period of about 1.2 ps. One also notices an extra modulation (around 0.1\%)
with a short pseudoperiod of about 30 fs and even shorter convoluted temporal
structures.

These typical times can be rationalized in terms of
the eigenstate decomposition of the initial datum,
which corresponds almost perfectly (with 49$\%$) to a one-to-one
mixture of the even vs.\ odd members of the ground-state tunneling pair. As a
result, 
the initial wavepacket is localized on the ``reactant'' side. The
corresponding eigenenergies are 975.92 cm$^{-1}$ and 976.30 cm$^{-1}$
(wavenumbers will be used as customary energy equivalents within this
vibrational context). The tunneling energy splitting of 0.38 cm$^{-1}$
corresponds to a ground-state tunneling period of 88.0 ps, as
indeed observed and reported above (long time scale). Note that
the single-well harmonic approximation of the zero-point energy around the
origin is at 1010 cm$^{-1}$ (the first tunneling pair is redshifted by about
24 cm$^{-1}$ due to the anharmonicity of the double well). The initial
wavepacket  also contains to some extent
a contribution of the next tunneling pair with
respect to $X$: i.e., a 1$\%$ component of both
members of the excited-state tunneling pair, at 2728 cm$^{-1}$ and 2757
cm$^{-1}$, split by 29 cm$^{-1}$.  This induces a medium time
scale pertaining to the excited tunneling period of 1.2 ps, as
indeed observed. The shorter times are more subtle to interpret. The harmonic
approximation around the origin (with an energy quantum of 2000 cm$^{-1}$) would induce
a harmonic period of 17 fs. The actual difference between the average
eigenergies of the first two tunneling pairs is a bit lower, at 1767
cm$^{-1}$ with a time of 19 fs, on the same order of magnitude as what we
identified as a short pseudoperiod of 30 fs, which can be termed a
``quasiharmonic" time. The overall dynamics thus appears to be governed by a
four-level eigensystem organized as a ``pair of pairs". Note that the harmonic
period for $Y$ is 1.7 ps (with an energy quantum of 20 cm$^{-1}$), hence, slightly
larger than the medium timescale.

Apart from the details of the interpretation, the present 
setting is ideal for our study, as we are dealing with three typical
timescales that are well separated from each other by about two orders of
magnitude.

\begin{figure}
{\includegraphics[width = 0.49\textwidth]{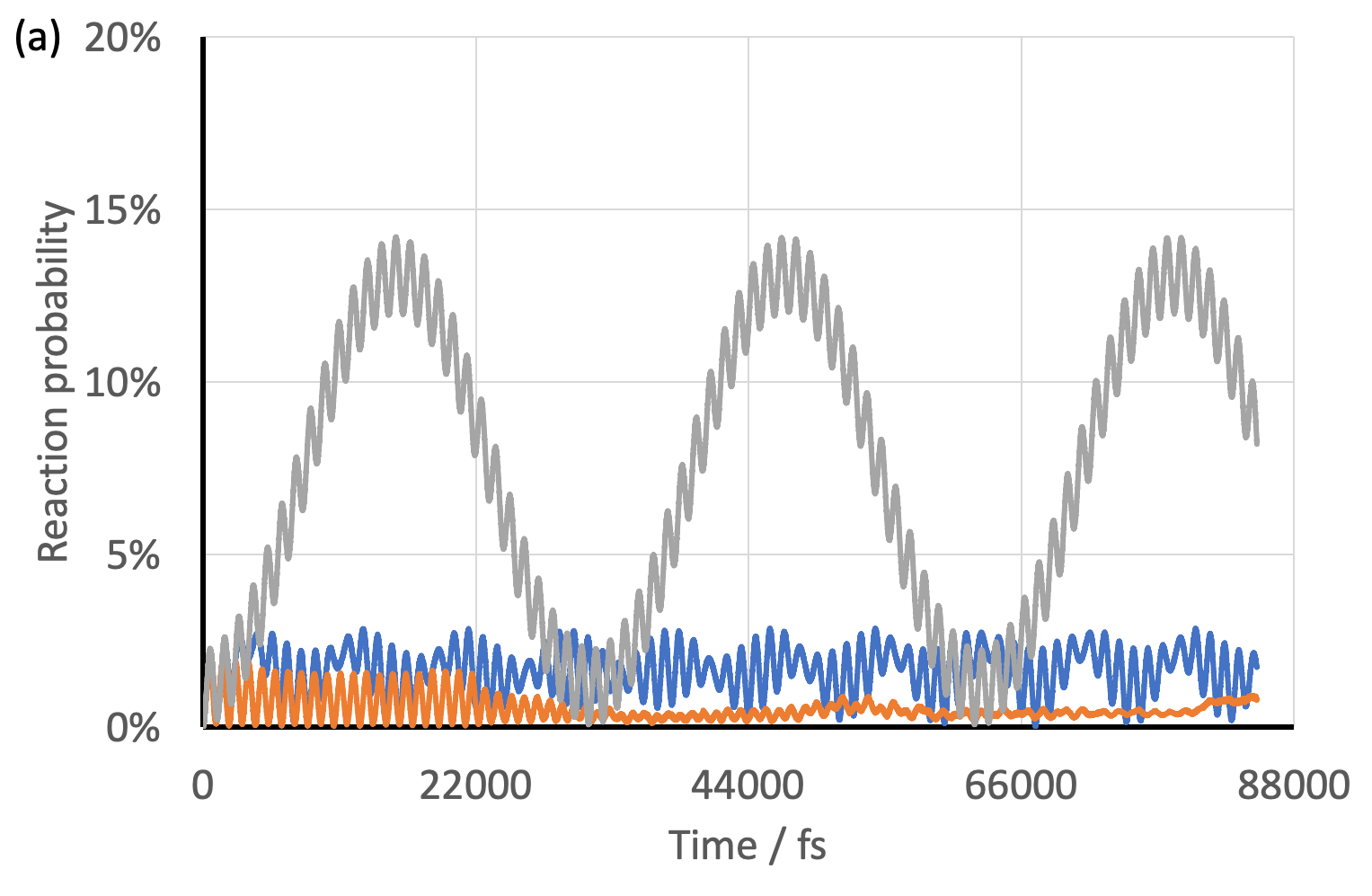}}
{\includegraphics[width = 0.49\textwidth]{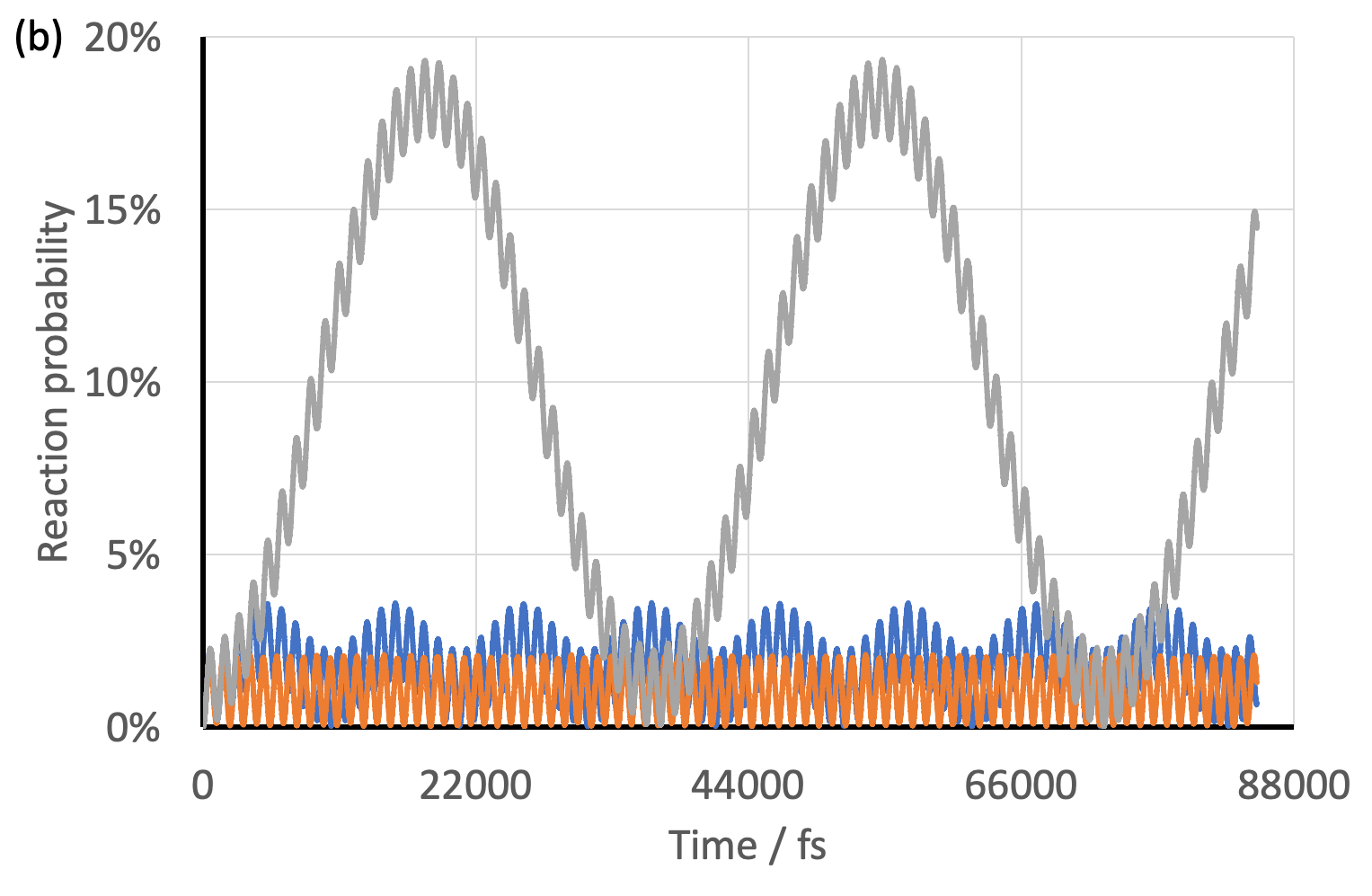}}\\
{\includegraphics[width = 0.49\textwidth]{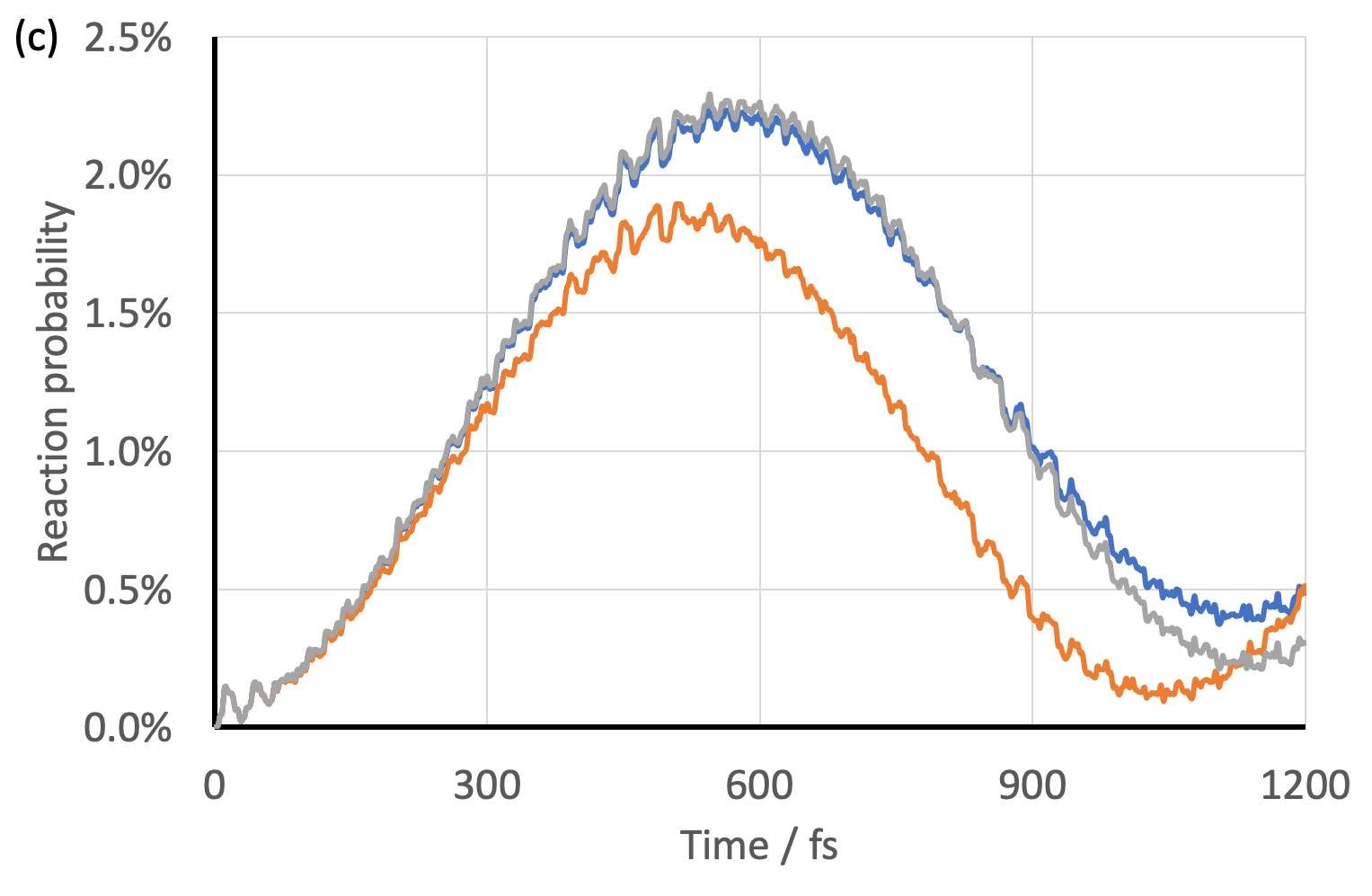}}
{\includegraphics[width = 0.49\textwidth]{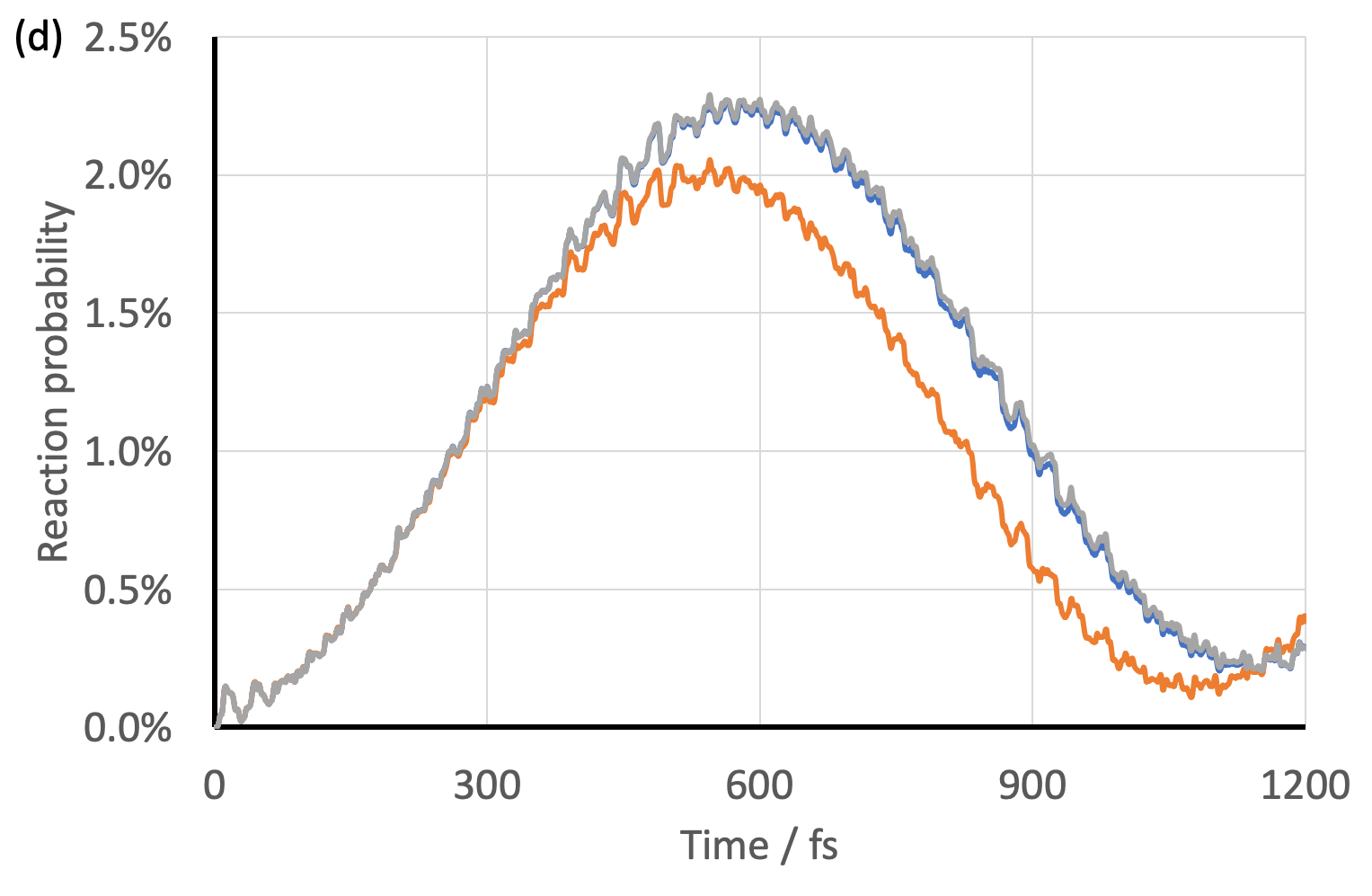}}
\caption{Fully correlated (MCTDH) vs.\ TDH propagation: Time evolution of
the reaction probability, $R(T)$, in three coupled cases (blue: case *;
orange: case 3; grey: case 4 -- see Table \ref{tab:cases}); the two left
panels show fully correlated (MCTDH), the two right panels TDH propagation; (a, b) long times
(ground-state tunneling); (c, d) medium times (excited-state tunneling).}
\label{fig:FigA2A3}
\end{figure}

\begin{figure}
{\includegraphics[width = 0.49\textwidth]{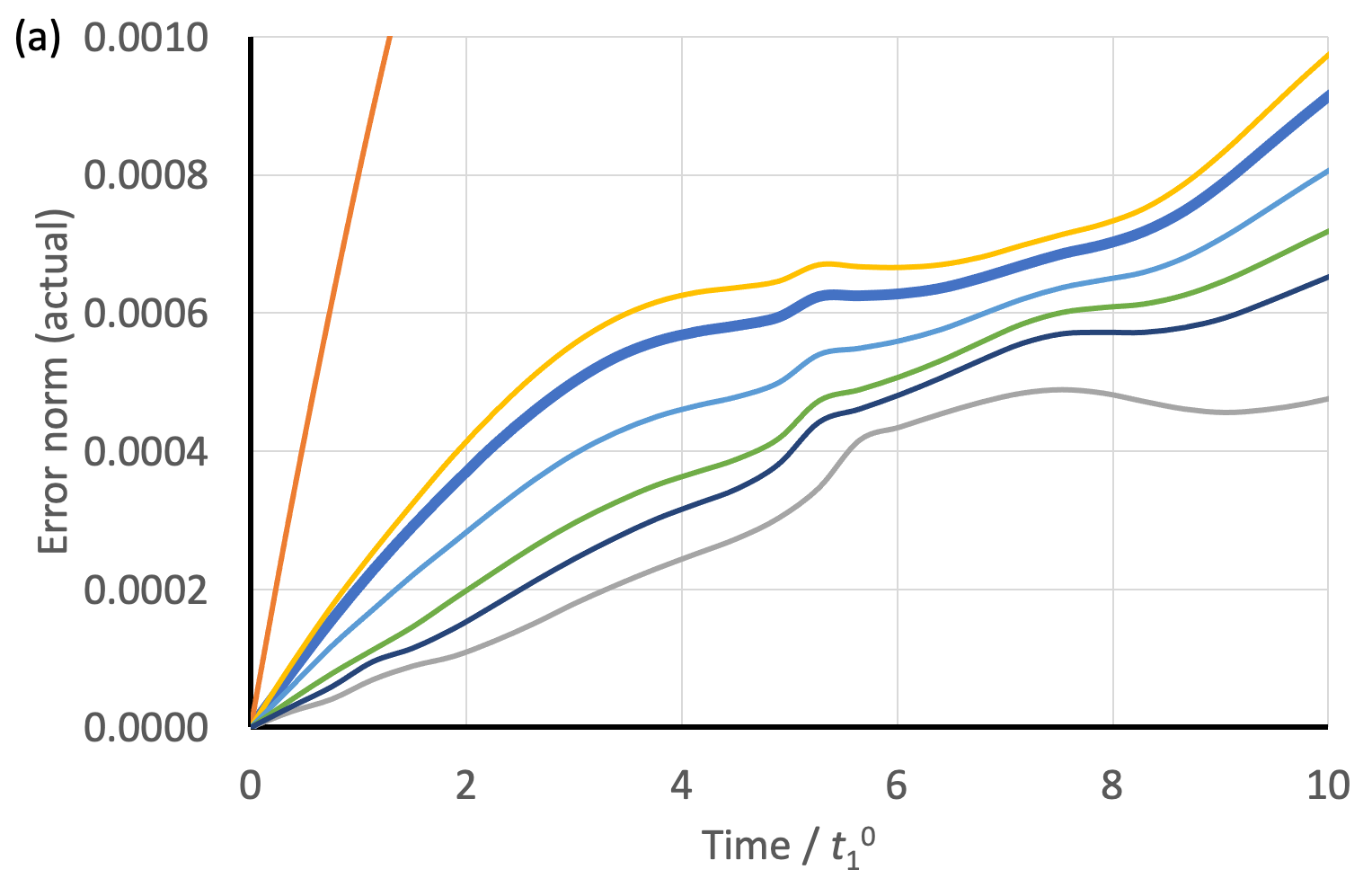}}
{\includegraphics[width = 0.49\textwidth]{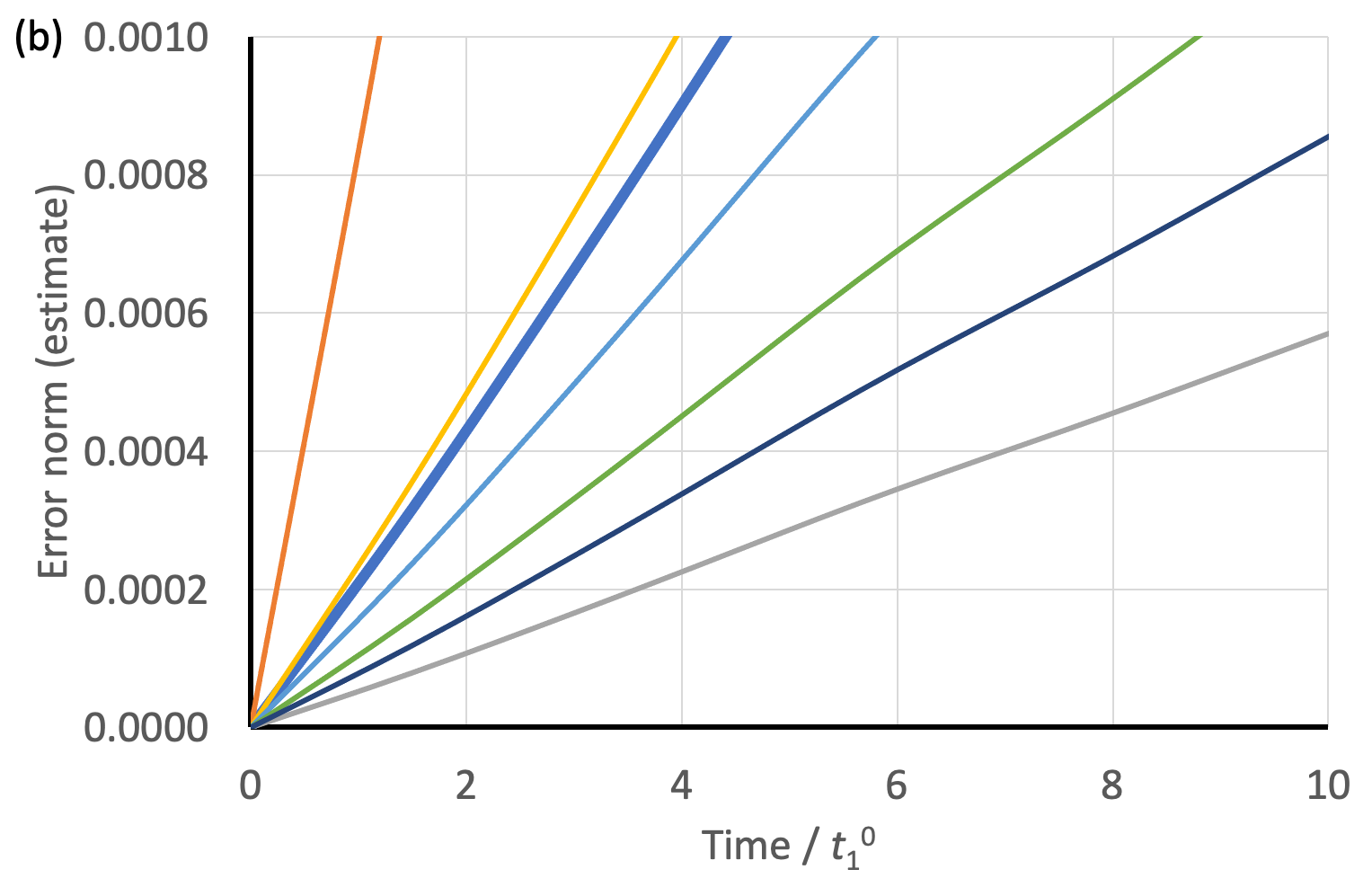}}\\
{\includegraphics[width = 0.49\textwidth]{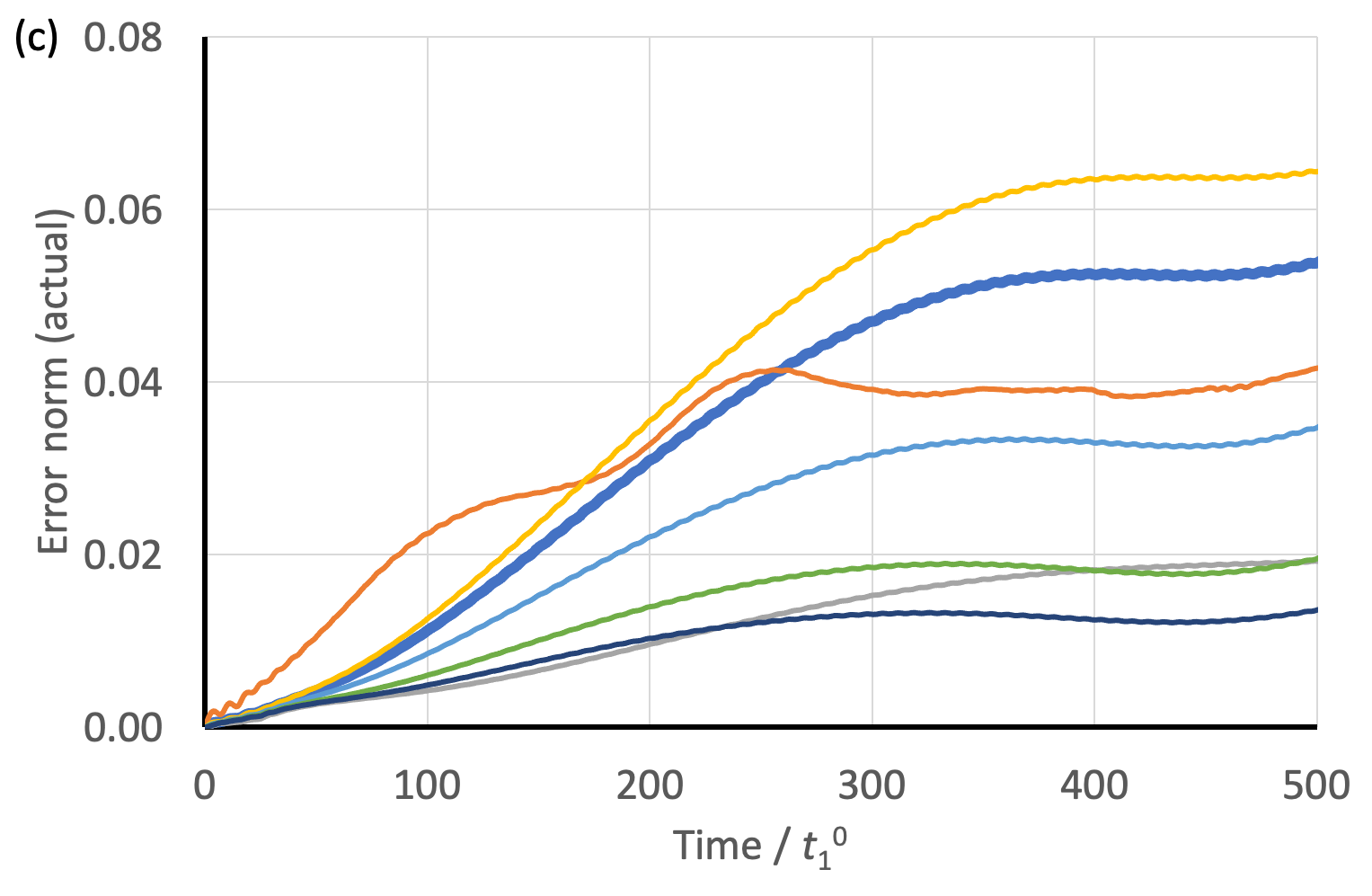}}
{\includegraphics[width = 0.49\textwidth]{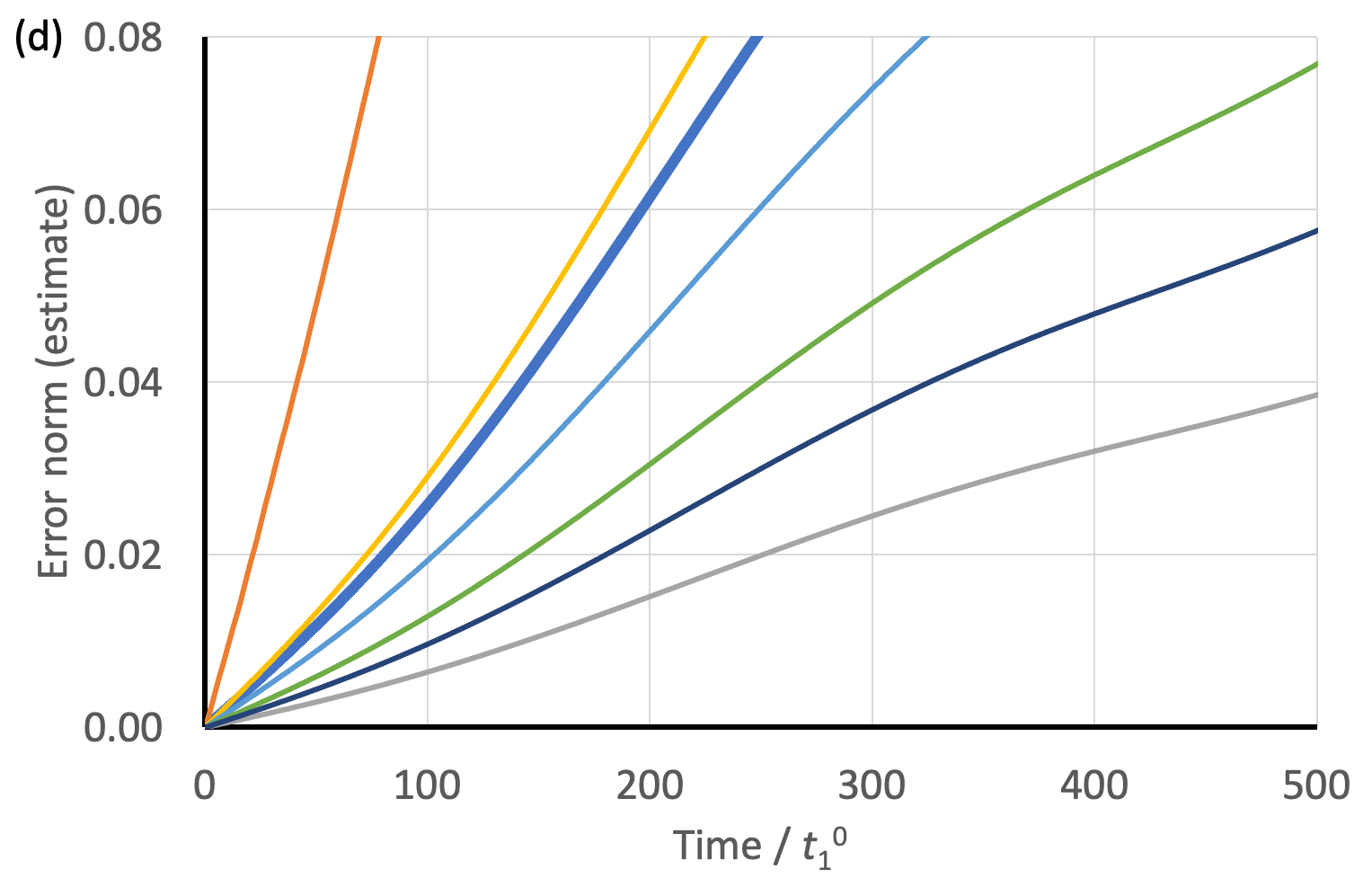}}\\
{\includegraphics[width = 0.49\textwidth]{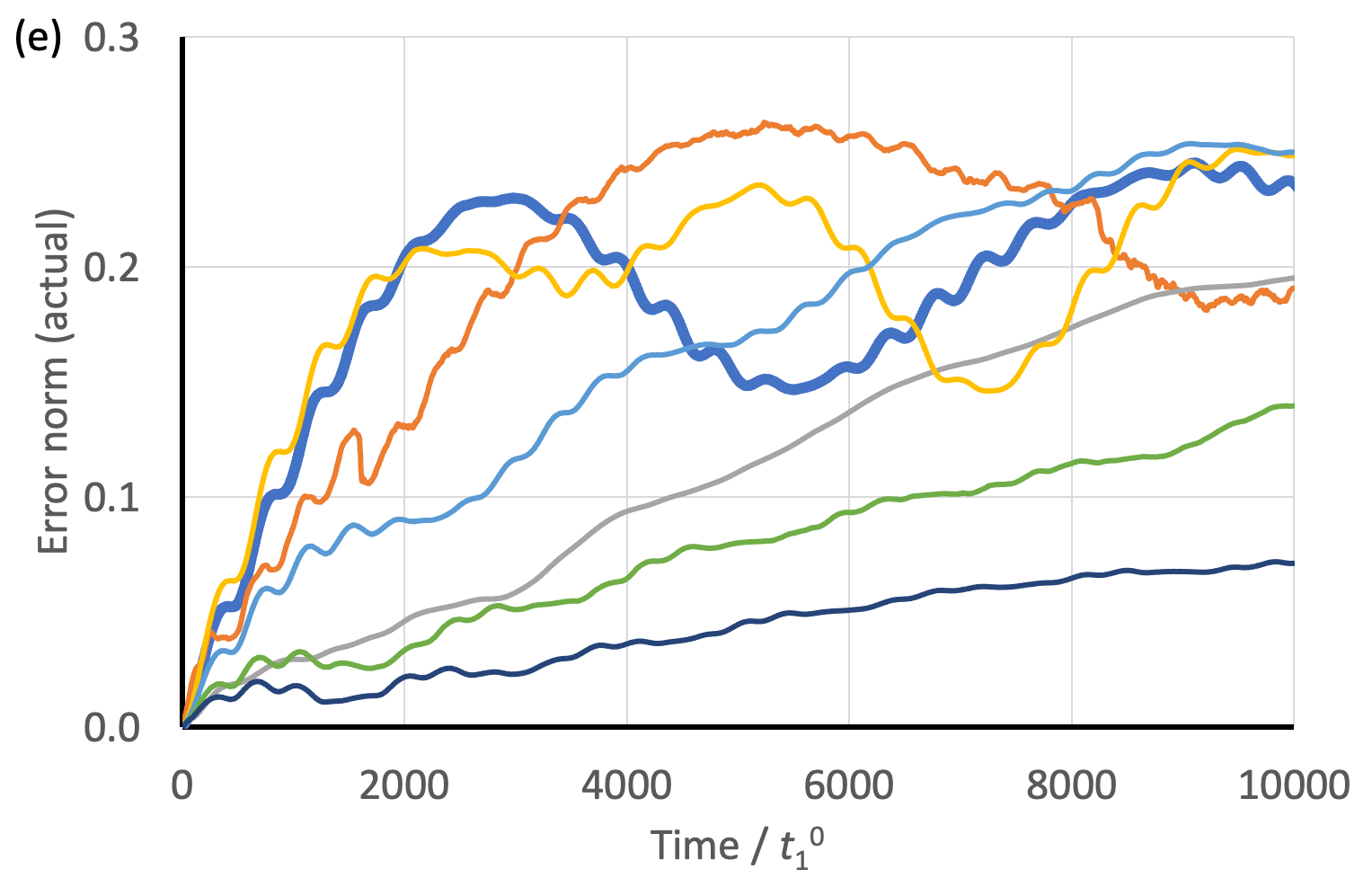}}
{\includegraphics[width = 0.49\textwidth]{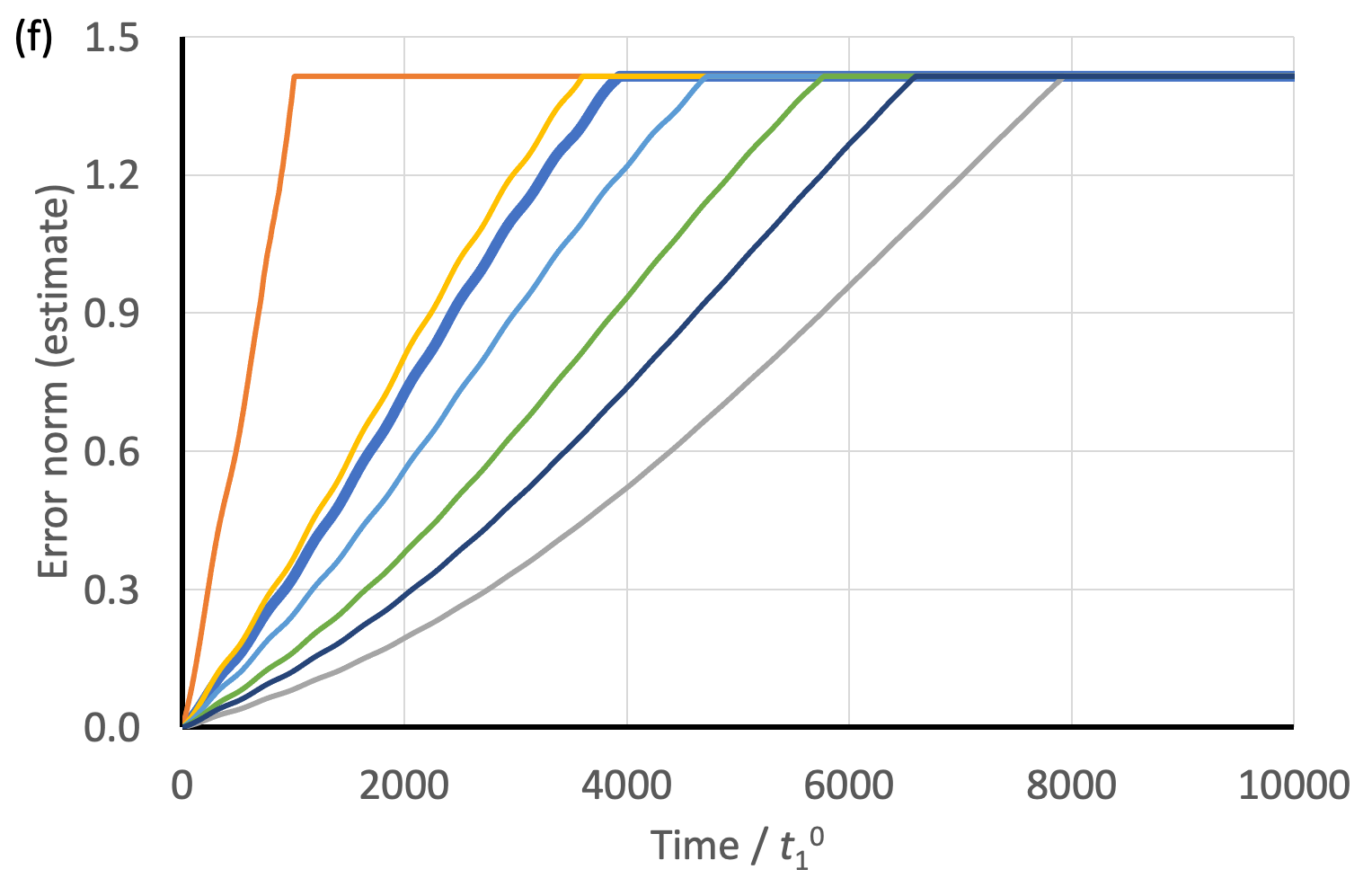}}
\caption{Time evolution of the norm $\|\Psi(T)-U(T)\|_{L^2}$ of the actual error and of the error estimate 
-- see Eq.~(\ref{eq:error-estimate}) -- in seven coupled cases (thick blue: case *; orange: case 3; grey: case 4; yellow: case 5; light blue: case 6; green: case 7;  dark blue: case 8 -- see Table \ref{tab:cases}); the three left panels show the actual error, the three right ones the estimate; (a, b) short times; (c, d) medium times; (e, f) long times.}
\label{fig:FigB1B2}
\end{figure}

\begin{figure}
{\includegraphics[width = 0.49\textwidth]{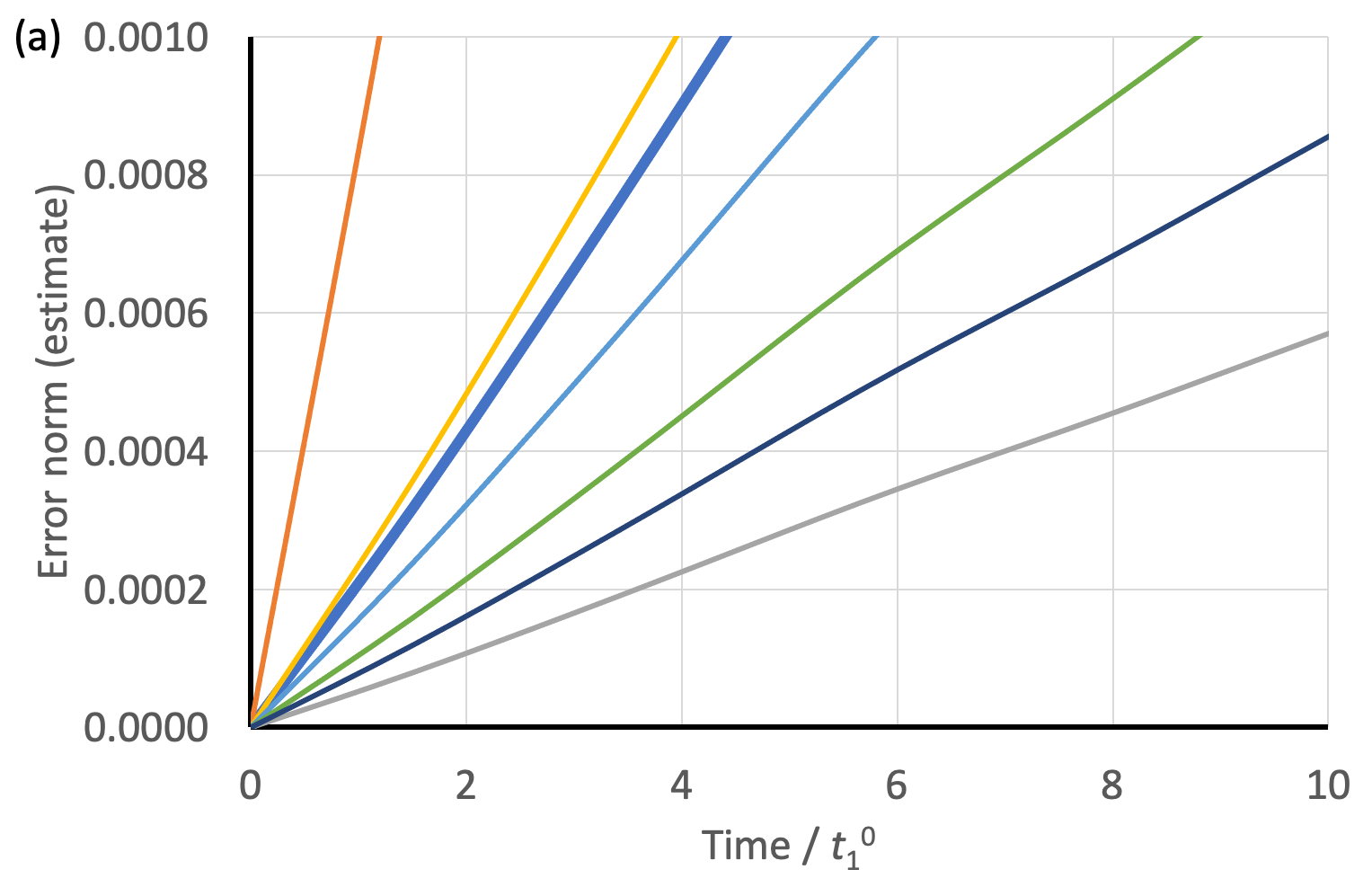}}
{\includegraphics[width = 0.49\textwidth]{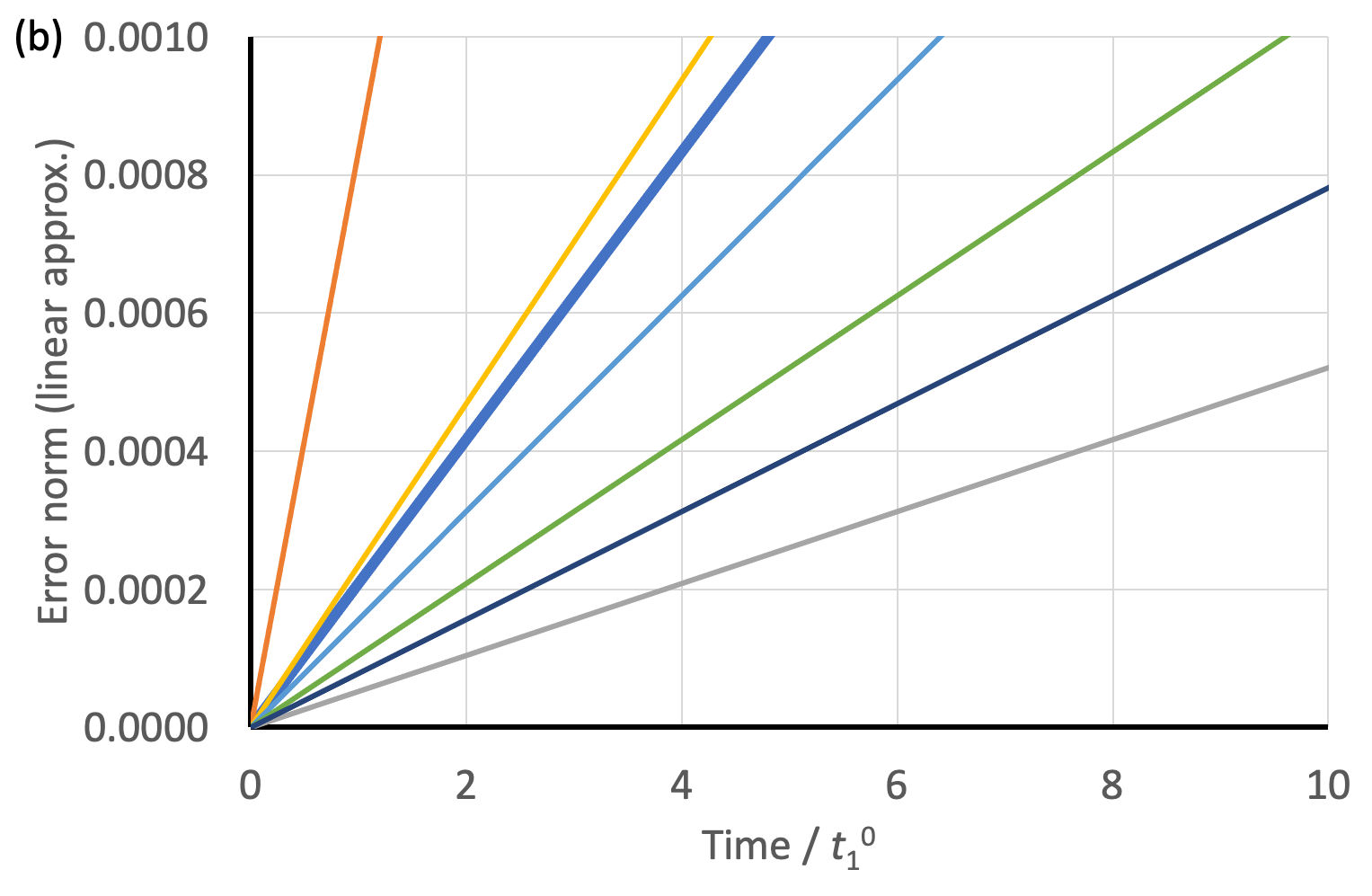}}\\
{\includegraphics[width = 0.49\textwidth]{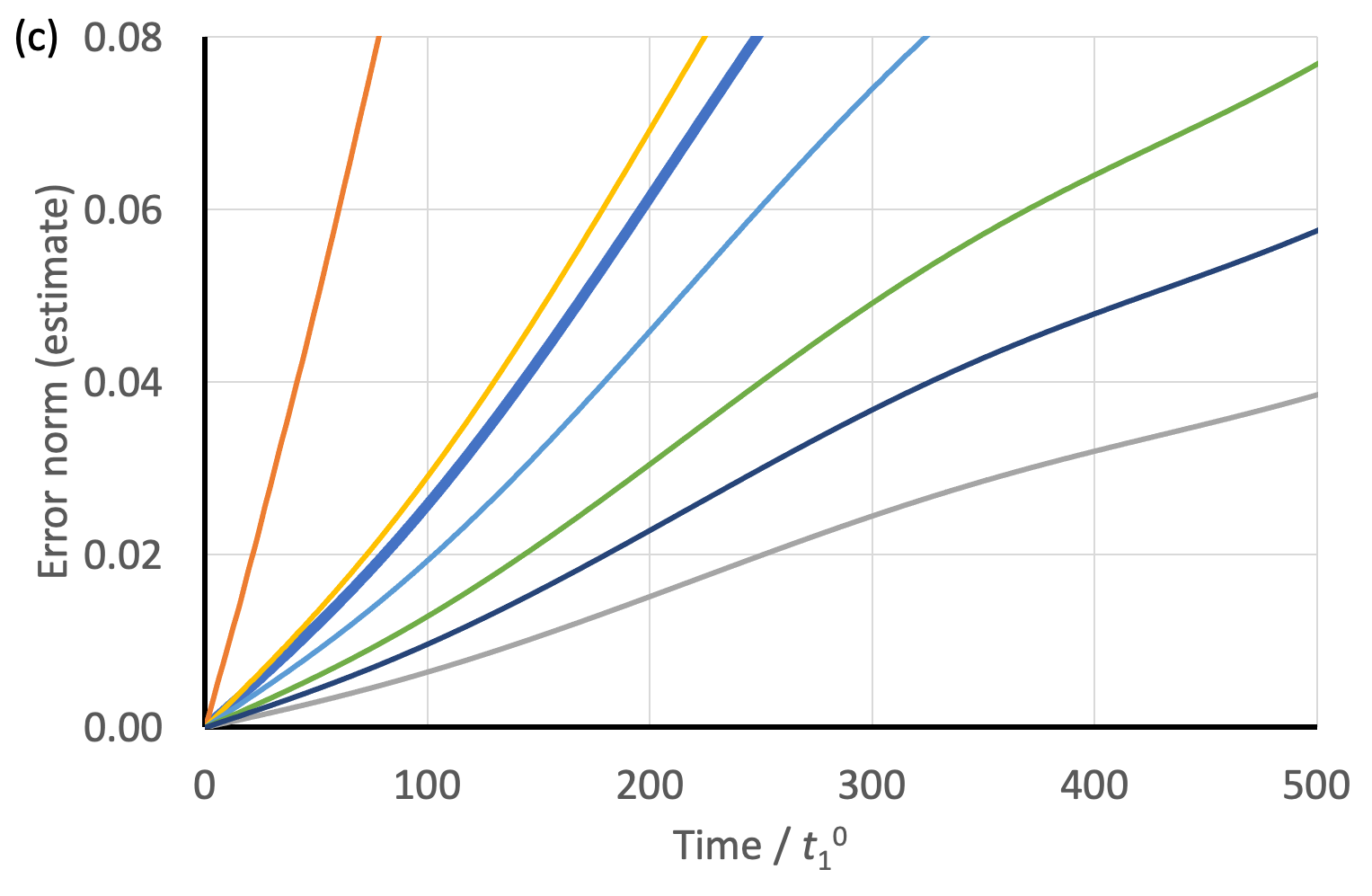}}
{\includegraphics[width = 0.49\textwidth]{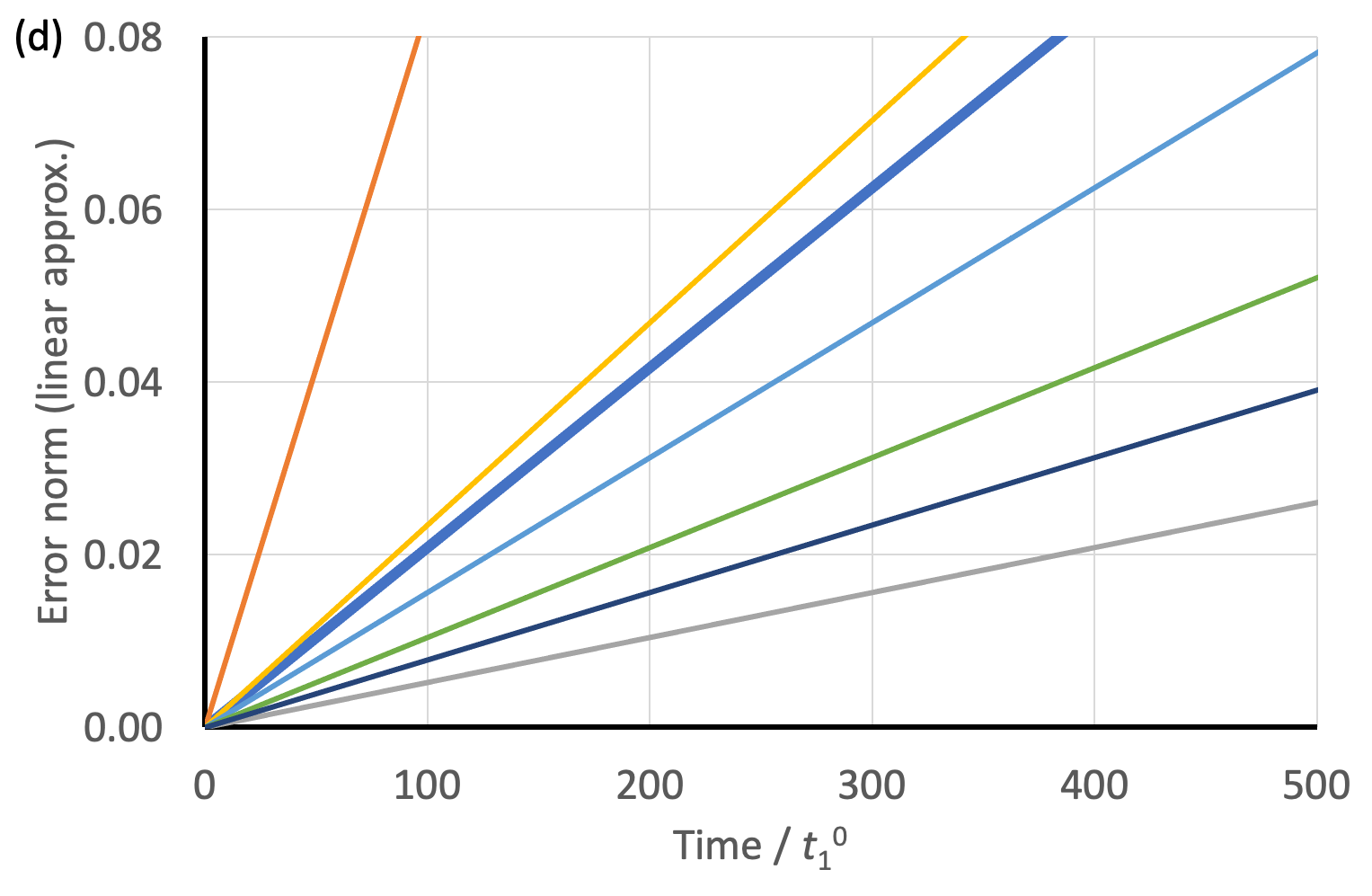}}\\
{\includegraphics[width = 0.49\textwidth]{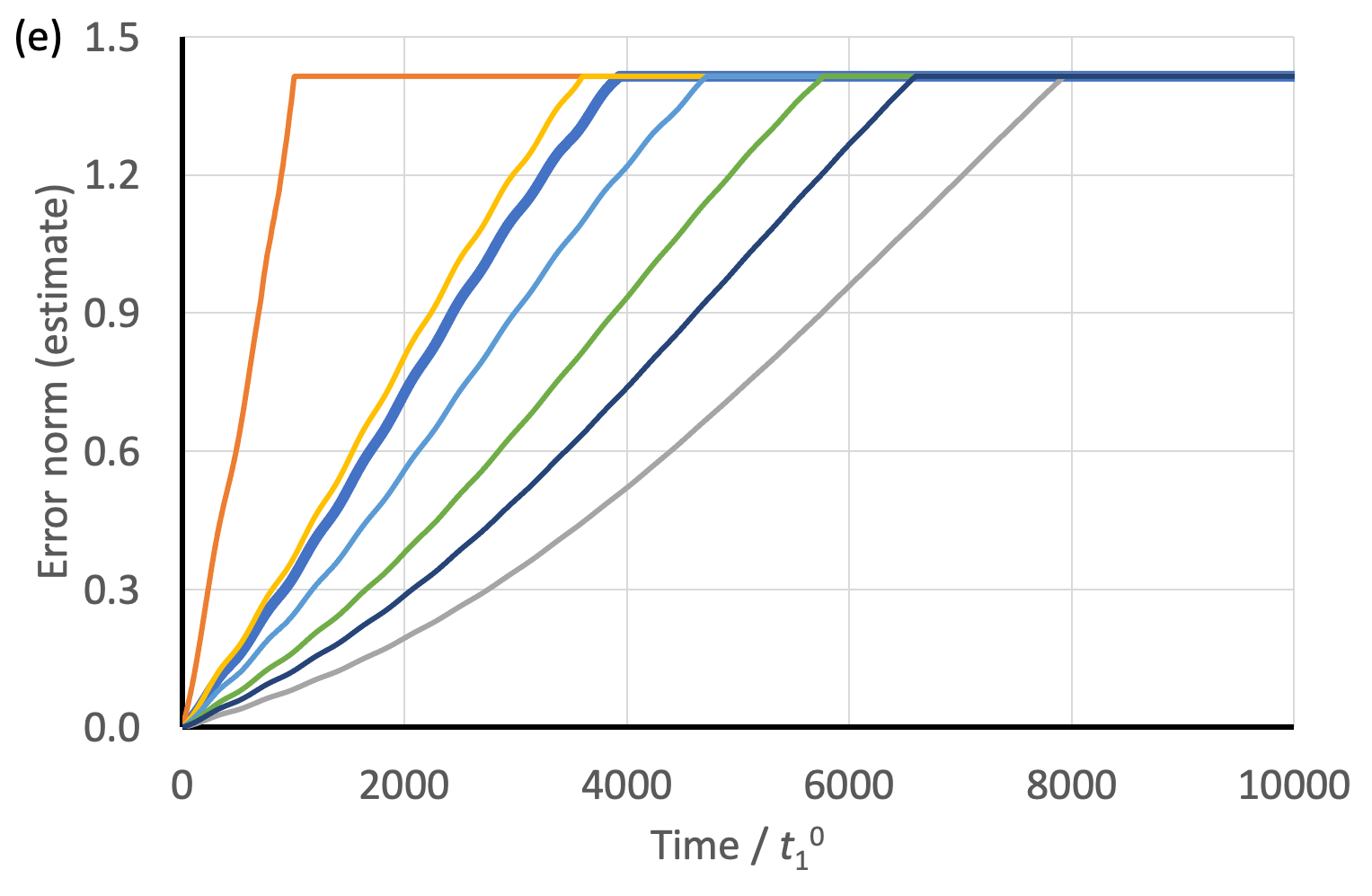}}
{\includegraphics[width = 0.49\textwidth]{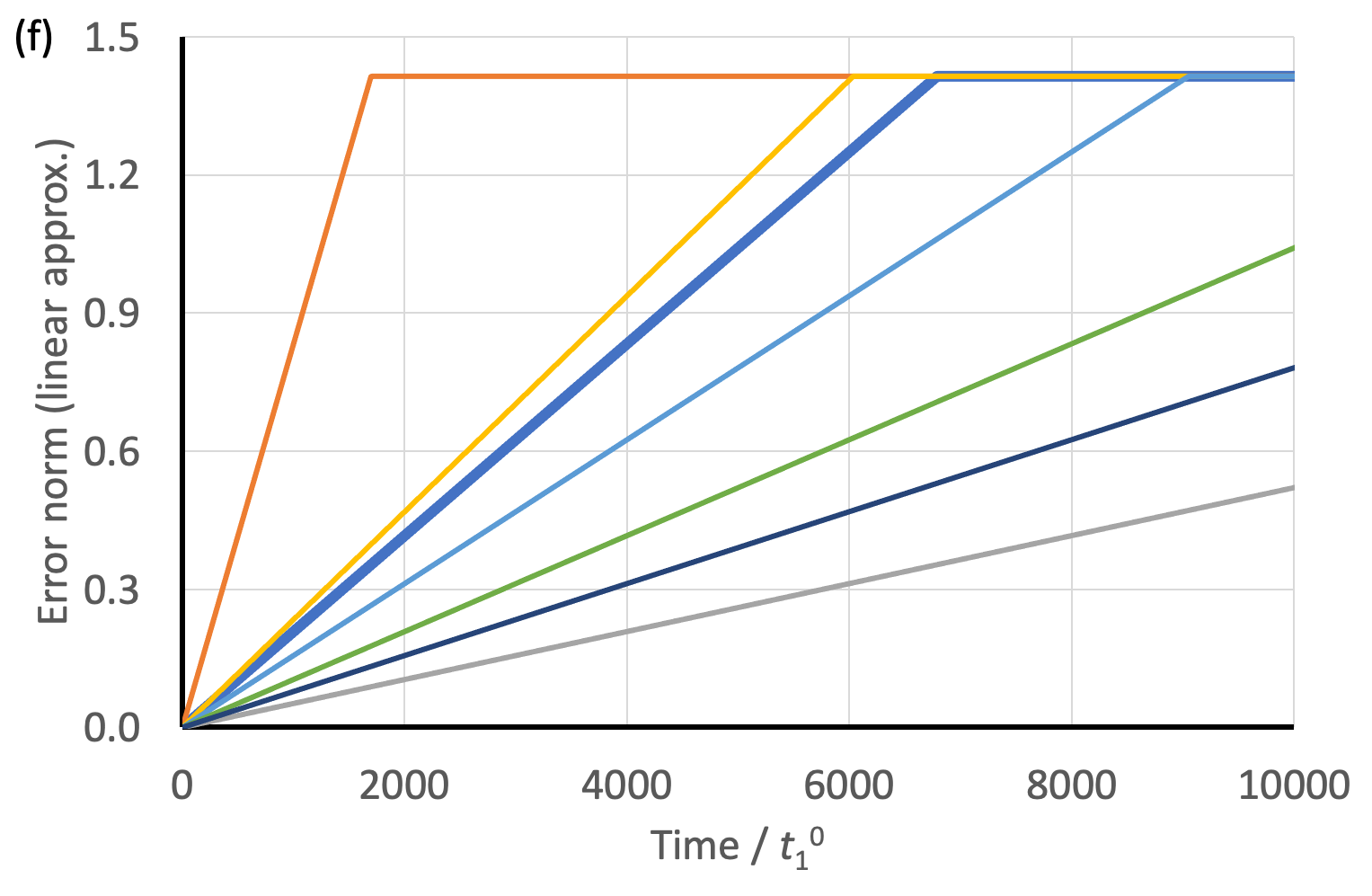}}
\caption{Time evolution of the error estimate -- see Eq.~(\ref{eq:error-estimate}) -- and the linear approximation -- see Eq.~(\ref{eq:linear-approx}) -- in seven coupled cases (thick blue: case *; orange: case 3; grey: case 4; yellow: case 5; light blue: case 6; green: case 7;  dark blue: case 8 -- see Table \ref{tab:cases}); the three left panels show the error estimate, the three right ones the linearization; (a, b) short times; (c, d) medium times; (e, f) long times.}
\label{fig:FigB2B3}
\end{figure}

We calculated the reaction probability for all cases given in
Table~\ref{tab:cases}. Results for the reference model ($*$) and
cases~3 and~4 are 
presented in Fig.~\ref{fig:FigA2A3}, for long and medium times both with the
fully correlated (MCTDH) and TDH methods. Short-time results are not shown, since
these are
all virtually identical to the uncoupled case (the dynamics is still
uncorrelated).

Upon comparing the left and right panels in Fig.~\ref{fig:FigA2A3}, we observe
significant differences between fully correlated (MCTDH) and TDH results at long times: (i)
the reaction probabilities obtained with TDH do not experience any oscillation damping
as opposed  to the fully correlated calculations; (ii) {the TDH population transfer
is slower (smaller rate); (iii) the net transfer is bigger (larger yield).

From a local comparison in time, the error could be considered as
very large.
However, from a global perspective, the dynamical behavior is qualitatively
similar and the orders of magnitude are correct. For example, taking case 4
(grey curves), the tunneling period is 31 ps (fully correlated) \emph{vs.} 37 ps
(TDH) and the rate is 14\% (fully correlated) \emph{vs.} 19\% (TDH). Let us now turn
to a more detailed analysis of the error  in 
 the time-dependent
wavefunction.

\subsection{Numerical assessment of error estimates}

We calculated the actual norm of the error between fully
correlated (MCTDH) and TDH
wavepackets at all times, $\|\Psi(T)-U(T)\|_{L^2}$, in the seven coupled cases
given in Table~\ref{tab:cases} (let us remind that cases 1 and 2 are identical to case * up to within homothetic dynamics in $Y$). These are shown in the left panels of
Fig.~\ref{fig:FigB1B2}. We also calculated the error estimate given in Eq.~(\ref{eq:error-estimate}), as well as its linear approximation defined in Eq.~(\ref{eq:linear-approx}), see Fig.~\ref{fig:FigB2B3}.

The error estimate given in Eq.~(\ref{eq:error-estimate}) appears to be almost exact over dimensionless times $\sim$ 100 $t^0_1$ where $t^0_1 =$ 2.65 fs (natural time for $X$). It is still dominated by a linear growth with time, as illustrated by its strong similarity with its linear approximation (Eq.~(\ref{eq:linear-approx})). The latter stays quite identical to the former up to about 100 $t^0_1$ (compare both panels (b) in Figs \ref{fig:FigB2B3}). This reflects small variations of the various moments, consistent with using a quasi-coherent state as initial datum.

Both rigorous and approximate error estimates keep cases ordered over time (no crossing between curves) according to the value of the linear prefactor, $\varpi \varsigma$, while the actual error starts to become more complicated, showing various types of oscillations and some rough saturation around 0.2 to 0.3.

Our estimates keep increasing and stop becoming relevant at $\sim$ 1000 $t^0_1$;
they still can be viewed as upper bounds, though. Note that they finally lose
any significance when they reach the critical value $\sqrt{2}$, where
orthogonality between the approximate and exact solutions sets in.

\section{Conclusions and outlook}

Following up on the recently presented mathematical framework for error
estimation in the context of composite quantum systems \cite{BCFLL}, we
presented here a first application to a non-trivial two-dimensional system
where tunneling motion (i.e., a ``reactive subsystem'') is coupled to a
quasi-harmonic degree of freedom via a cubic coupling. The
values of our system Hamiltonian parameters (frequencies, masses, tunneling
length and barrier; see Supplementary Material [LINK]) were chosen so as to
correspond to realistic molecular situations. As it occurs, they allow for
significant quantum effects, in particular an interesting tunneling process
that exhibits three distinct characteristic time scales.

For reference, the reliability of a separable
time-dependent Hartree ansatz for the time-evolving wavepacket was assessed by 
comparison with converged multiconfigurational (MCTDH) calculations that
can be considered as numerically exact. The relevant space of system
parameters was explored with respect to the coupling strength as well as the
relative timescales of the subsystem and the bath.

In the parameter regimes we considered, the TDH approximation
represents a good zeroth-order approximation
which requires corrections such as to account for correlations. In line with this,
%\textcolor{red}{decent approximation (say, a good zero-order starting point, qualitatively, which shows room for quantitative improvement according %to a hierarchy of better approximations dealing explicitly with correlation effects)}, i.e., the dynamical evolution is reproduced
%in a qualitatively correct way,
tunneling rates and yields differ by
no more than a factor of two from the exact result. Yet, the quantitative
error is non-negligible, such that the error estimates developed in
Ref.\ \cite{BCFLL} are relevant. In the present study, this error is
numerically computed and compared with our rigorous
mathematical estimates \cite{BCFLL}. These estimates were shown to provide
a good approximation to the numerically exact error, and yield an
almost exact result in the early regime of near-linear growth of the error with time before
saturation.

Against the background of a detailed scaling analysis, we
further introduced a linearization approach by which an expression for the
short-time error estimate was derived, which is found to depend on the
frequency ratio of the subsystems. This emphasizes that from the
vantage point of the TDH approximation, the frequency ratio rather than
the mass ratio of the subsystems is of crucial importance.
Again, the linearization estimate was found to provide a valid approximation,
even beyond the shortest time scale.

The present work paves the way for
extensions of error analysis to other types of wavefunction {\em ansatz},
such as multiconfigurational forms of MCTDH type
\cite{BECK20001,GattiLasorne}, and especially Gaussian-based hybrid
wavefunctions such as employed in the 
G-MCTDH method \cite{Burghardt,RoeBurg}. The model that we used showed moderate
failures of TDH that have observable consequences on, for example, the
reaction probability. It could thus be useful for benchmarking a hierarchy of
methods of various sophistication.

Finally, as mentioned in Sec.\ 2, the present treatment can be
generalized to a genuine system-bath situation where the subsystems are
subject to external fluctuations inducing dissipation \cite{CL:81,Leggett:87}.
In this context, the present perspective immediately connects to a
non-Markovian treatment of structured environments which can be decomposed in
terms of effective environmental modes \cite{Martinazzo:11,Martinazzo2:11}.
From this viewpoint, the system-bath boundary can be shifted such as to
include a set of environmental modes as additional subsystems in an explicit
treatment, while a residual bath is included by a Markovian approximation.
This type of approach has been recently employed in the context of
two-dimensional molecular tunneling dynamics \cite{Picconi19}, in a similar
parameter range as specified in the present model. These models permit to
further investigate fluctuation-induced enhancement or reduction of tunneling,
localization effects, and decoherence, which are ubiquitous effects far beyond
the molecular tunneling situation considered here. Including fluctuations and
dissipation is obviously of general importance in quantum metastable systems
\cite{Weiss:12,Breuer:02,Spagnolo2:18}. This direction provides a
natural extension to the treatment employed in the present work.

\section*{Acknowledgements.}  Adhering to the prevalent convention in mathematics, we chose an alphabetical ordering of authors. We warmly thank Lucien Dupuy, G\'erard Parlant, Yohann Scribano, and Graham A. Worth for important discussions and support during the genesis and redaction of this article. We thank the CIRM for twice hosting our interdisciplinary team and the CNRS 80 $\vert$ Prime project for support.

%%%%%%%%%% Insert bibliography here %%%%%%%%%%%%%%

\bibliographystyle{unsrt}
\bibliography{biblio}

%%%%%%%%%% APPENDIX %%%%%%%%%%%%%%%%%%%%%%%%%%%

\end{document}